\documentclass[11pt,a4paper,twoside,openright]{article}
\usepackage{latexsym}
\usepackage{graphics}
\usepackage{float}
\usepackage[utf8]{inputenc}
\usepackage[english]{babel}
\usepackage[pdftex]{graphicx}
\usepackage[font=small,format=plain,labelfont=bf,up,textfont=it,up]{caption}
\usepackage{slashed}
\usepackage{bm}
\usepackage{bbm}
\usepackage{fancyhdr}
\fancyheadoffset{0cm}
\fancyfootoffset{0cm}
\usepackage{amssymb}
\usepackage{amsopn}
\usepackage{amsmath} 
\usepackage{varioref}

\usepackage{amssymb}
\usepackage{amsopn}
\usepackage{amsmath}
\usepackage{bbm}
\if@mathematic
   \def\vec#1{\ensuremath{\mathchoice
                     {\mbox{\boldmath$\displaystyle\mathbf{#1}$}}
                     {\mbox{\boldmath$\textstyle\mathbf{#1}$}}
                     {\mbox{\boldmath$\scriptstyle\mathbf{#1}$}}
                     {\mbox{\boldmath$\scriptscriptstyle\mathbf{#1}$}}}}
\else
   \def\vec#1{\ensuremath{\mathchoice
                     {\mbox{\boldmath$\displaystyle#1$}}
                     {\mbox{\boldmath$\textstyle#1$}}
                     {\mbox{\boldmath$\scriptstyle#1$}}
                     {\mbox{\boldmath$\scriptscriptstyle#1$}}}}
\fi

\makeatletter

\renewcommand\section{\@startsection {section}{-1}{\z@}%
                                   {-3.5ex \@plus -1ex \@minus -.2ex}%
                                   {2.3ex \@plus.2ex}%
                                   {\bfseries \sffamily \large }}

\renewcommand\subsection{\@startsection {subsection}{1}{\z@}%
                                   {-3.5ex \@plus -1ex \@minus -.2ex}%
                                   {2.3ex \@plus.2ex}%
                                   {\bfseries \sffamily\large}}
\makeatother

\usepackage{mathrsfs}
\usepackage{lmodern}
\usepackage{url}
\usepackage{xcolor}
\definecolor{dark}{rgb}{0.10,0.2,0.3}
\definecolor{light}{rgb}{1.7,1.5,0.6}
\definecolor{purpure}{rgb}{0.5,0.15,0.3}
\usepackage{hyperref}
\hypersetup{colorlinks,%
citecolor=dark,%
filecolor=dark,%
linkcolor=purpure,%
urlcolor=purpure,%
pdftex}

\usepackage{graphicx}
\begin{document}

\renewcommand{\headrulewidth}{0.4pt}
\renewcommand{\footrulewidth}{0pt}

\renewcommand{\contentsname}{\hfill\LARGE\bfseries \sffamily Table of Contents}
\renewcommand{\bibname}{\hfill\LARGE\bfseries \sffamily References}
\newpage
\pagestyle{plain}
\title{ \LARGE   \bf  NLO corrections to the gluon induced forward jet vertex 
from the high energy effective action}
\date{}
\maketitle 
\vspace{-1.5cm}
\begin{center}
  { \large   G.~Chachamis$^1$, M.~Hentschinski$^2$, J.~D.~Madrigal Mart\'inez$^{3,4}$, A.~Sabio Vera$^{3,4}$}
\end{center}
\vspace{.2cm}
\begin{verse}
{ $^1$~Instituto de F{\' \i}sica Corpuscular UVEG/CSIC,
46980 Paterna (Valencia), Spain}.\\ 
{ $^2$~Department of Physics, Brookhaven National Laboratory, Upton, NY 11973,
USA.}\\ 
{ $^3$~Instituto de F\'isica Te\'orica UAM/CSIC, Nicol\'as Cabrera 15,  28049 Madrid, Spain.}\\ 
{ $^4$~Facultad de Ciencias, Universidad Aut\'onoma de Madrid, C.U. Cantoblanco,  \\ $\,\,$ 28049 Madrid, Spain.}
\end{verse}

\vspace{0.5cm}
\begin{abstract}
  We determine both real and virtual next-to-leading order corrections to
  the gluon induced forward jet vertex, from the high energy effective
  action proposed by Lipatov. For these calculations we employ the same 
  regularization and subtraction formalism developed in our previous work on the quark-initiated 
  vertex.  We find agreement with previous results in the literature.
\end{abstract}
 {\small   }

\vspace{0.4cm}

\section{Theoretical framework}
\label{1}

In this work we present the calculation of the vertex describing the
production of a jet in a forward direction very close in the detector
to one of the hadrons in hadron-hadron interactions at very high
energies. This is done in the kinematic approximation where the jet
is well separated in rapidity from other jets also produced in the
scattering process. As a calculational technique we make use of
Lipatov's effective action~\cite{Lipatov:1995pn}, designed to easy the
derivation of scattering amplitudes in the high energy limit of
QCD. Here we focus on the gluon-initiated jet vertex, which is a more
complicated counterpart of the quark-initiated vertex, derived with
similar techniques in Ref.~\cite{Hentschinski:2011tz}. A convenient
regularization and subtraction procedure, taken from
Ref.~\cite{Hentschinski:2011tz}, is shown to give the correct results
(in agreement with previous calculations in the literature~\cite{Fadin:1993wh,Fadin:1992zt, Lipatov:1996ts, DelDuca:1998kx,Ciafaloni:1998hu,Fadin:1999de})
at next-to-leading (NLO) accuracy. The convolution of this jet vertex
with the NLO BFKL gluon Green function plays a very important role
in the description of jet production at the LHC physics program. Many interesting studies 
have been performed in this direction in recent years, see Refs.~\cite{Bartels:2001ge,Bartels:2002yj,Vera:2006un,Vera:2007kn,Marquet:2007xx,Vera:2007dr,Deak:2009xt,Colferai:2010wu,Caporale:2011cc,Ivanov:2012ms,Ivanov:2012iv,Caporale:2012ih}. \\
\\
A complete description of the high energy effective action used in
this work can be found in
Refs.~\cite{Lipatov:1995pn,Lipatov:1996ts}. For a more recent
discussion directly related to our calculation we refer the reader to
Ref.~\cite{Chachamis:2012mw}. The calculation of the quark contributions to the gluon
Regge trajectory at two loops using Lipatov's effective action has been performed by us in
Ref.~\cite{Chachamis:2012gh,Chachamis:2012tt}.  Here we will just
briefly explain the general structure of this action to then describe
in some detail our calculation of the gluon-initiated jet vertex. \\

\begin{figure}[htb]
  \centering
  \parbox{.4
    \textwidth}{\center \includegraphics[width=0.2\textwidth]{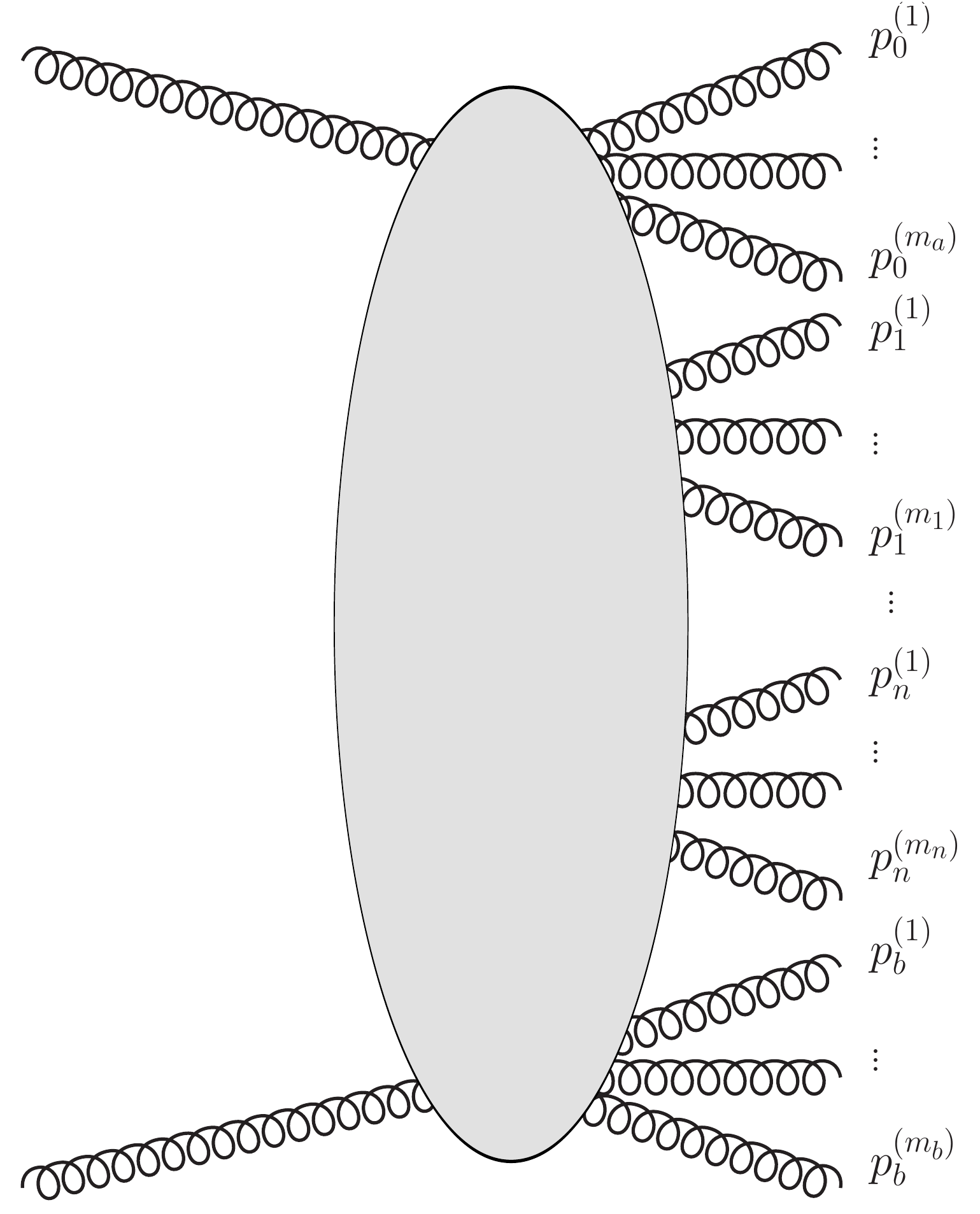}
  }
    \parbox{.5\textwidth}{  \center \includegraphics[width = .55\textwidth]{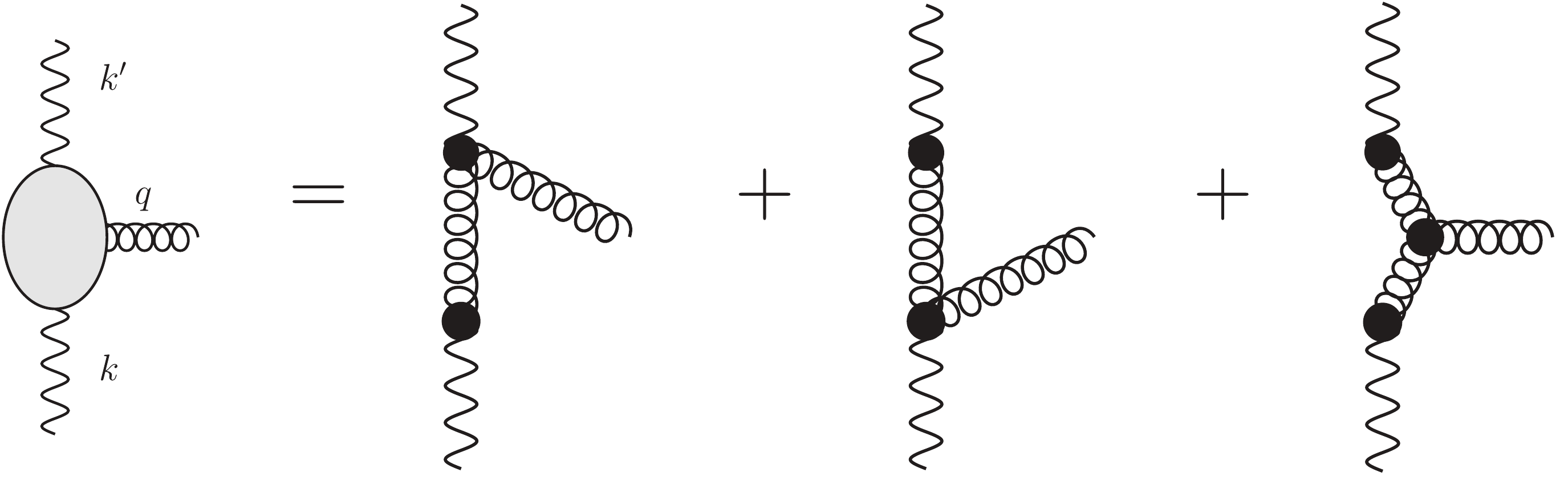}   }
        
 \parbox{.2 \textwidth}{ \center (a) } \parbox{.5 \textwidth}{ \center (b)} 
  \caption{\small   (a) Quasi-multi-regge kinematics; (b)  reggeized gluon-reggeized gluon-gluon effective vertex. The first two diagrams are the induced contributions.}
  \label{fig:1}
\end{figure}

The high energy effective action is based on the interplay between QCD
particles and reggeized degrees of freedom, which are introduced as
independent fields interacting with the standard ones via
new vertices.  These effective interactions, dominant in the
high energy limit of QCD, appear inside an extra term added to the QCD
action. Reggeized quarks and gluons ``propagate'' in the $t$-channel
with modified propagators connecting two regions of different
rapidities and play the role of suppressing any real emission in that
interval. At the endpoints of these intervals there can be particle production; single production in multi-regge kinematics (MRK) and
double production in quasi-multi-regge kinematics (QMRK). Within these
clusters of particle production there are not kinematic restrictions
and interactions are the usual QCD ones. A representation of this
effective clustering is shown in Fig.~\ref{fig:1}.a. The
ordering in rapidity of the produced clusters is of the form $y_0\gg
y_1\gg\cdots\gg y_{n+1}$, where $y_k=\frac{1}{2}\ln\frac{k^+}{k^-}$
and with all particles in each cluster being emitted with a similar
rapidity. Defining the light-cone vectors $n^{+,-}=2p_{a,b}/\sqrt{s}$
we work with Sudakov expansions for four dimensional vectors of the
form $k=\frac{1}{2}(k^+n^-+k^-n^+)+\vec{k}$, with ${\vec{k}}$ being
transverse.  The strong ordering of the clusters simplifies the
polarization tensor of the $t$-channel reggeized particles which can
be written as $g_{\mu\nu}\to\frac{1}{2}(n^+)_\mu (n^-)_\nu+{\cal
  O}(\frac{1}{s})$, with $\sqrt{s}$ being the center-of-mass energy of
the scattering process, carrying mainly transverse momenta,
$q_i^2=-\vec{q}_i^2$.

The effective interaction between reggeized and usual particles, like
the one shown in Fig.~\ref{fig:1}, consists of two pieces: the
projection of the QCD vertex onto high energy kinematics and
additional induced contributions. This structure can be obtained from
the following form of the effective action:
\begin{equation}\label{ea}
S_{\rm eff}=S_{\rm QCD}+S_{\rm ind};\quad S_{\rm ind}=\int d^4x\, {\rm Tr}[(W_+[v(x)]-\mathscr{A}_+(x))\partial_\perp^2\mathscr{A}_-(x)]+\{+\leftrightarrow -\},
\end{equation}
where $\mathscr{A}_\pm$ are the gauge-invariant reggeized gluon fields
which satisfy the following kinematic constraint
\begin{align}
  \label{eq:constraint}
  \partial_+\mathscr{A}_- (x)=0 = \partial_-\mathscr{A}_+ (x).
\end{align}
 Their couplings to the QCD gluon
field is given in terms of two non-local functionals $W_\pm
[v]=v_\pm\frac{1}{D_\pm}\partial_\pm=v_\pm-gv_\pm\frac{1}{\partial_\pm}v_\pm+\cdots$ with $D_\pm = \partial_\pm + g v_\pm $.
\begin{figure}[htb]
  \centering
   \parbox{.7cm}{\includegraphics[height = 1.8cm]{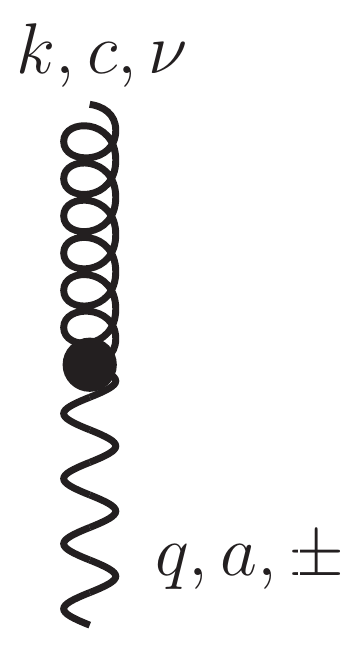}} $=  \displaystyle 
   \begin{array}[h]{ll}
    \\  \\ - i{\bm q}^2 \delta^{a c} (n^\pm)^\nu,  \\ \\  \qquad   k^\pm = 0.
   \end{array}  $ 
 \parbox{1.2cm}{ \includegraphics[height = 1.8cm]{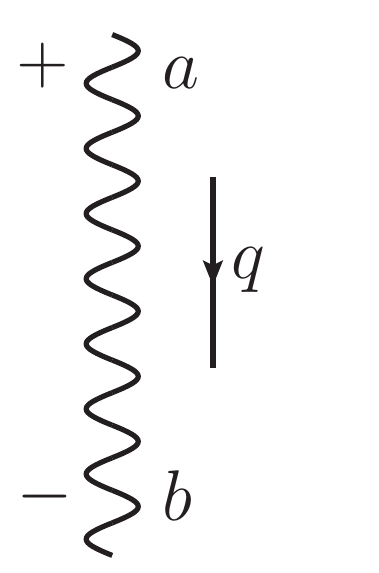}}  $=  \displaystyle    \begin{array}[h]{ll}
    \delta^{ab} \frac{ i/2}{{\bm q}^2} \end{array}$ 
 \parbox{1.7cm}{\includegraphics[height = 1.8cm]{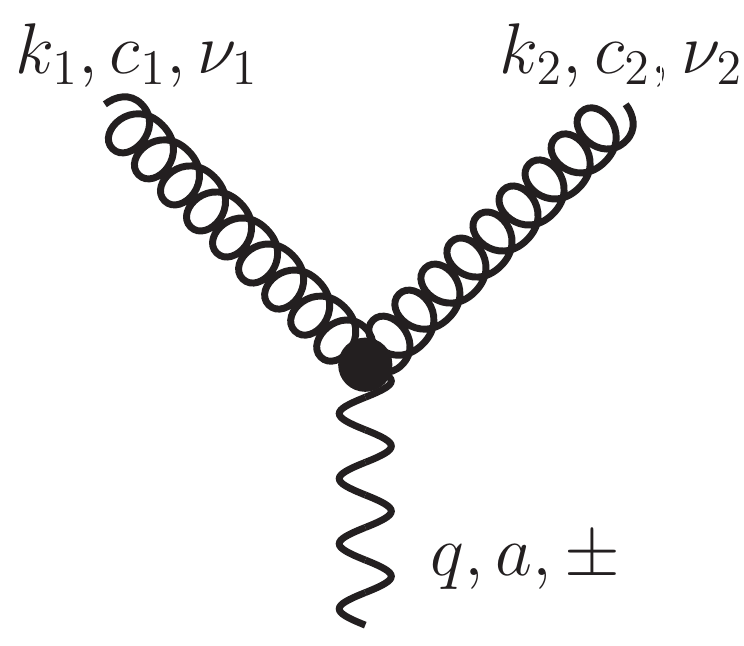}} $ \displaystyle  =  \begin{array}[h]{ll}  \\ \\ g f^{c_1 c_2 a} \frac{{\bm q}^2}{k_1^\pm}   (n^\pm)^{\nu_1} (n^\pm)^{\nu_2},  \\ \\ \quad  k_1^\pm  + k_2^\pm  = 0
 \end{array}$
 \\
\parbox{4cm}{\center (a)} \parbox{4cm}{\center (b)} \parbox{4cm}{\center (c)}

\vspace{1cm}
  \parbox{2.4cm}{\includegraphics[height = 1.8cm]{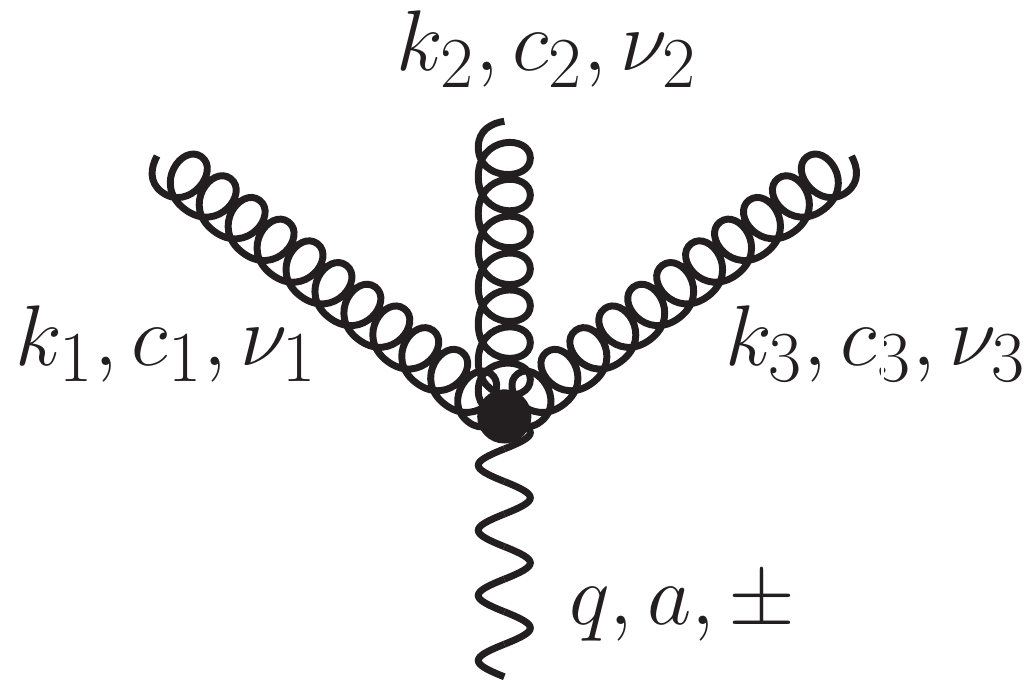}} $ \displaystyle 
   \begin{array}[h]{l}  \displaystyle  \\ \displaystyle= ig^2 {\bm{q}}^2 
\left(\frac{f^{a_3a_2 e} f^{a_1ea}}{k_3^\pm k_1^\pm} 
+
 \frac{f^{a_3a_1 e} f^{a_2ea}}{k_3^\pm k_2^\pm}\right) (n^\pm)^{\nu_1} (n^\pm)^{\nu_2} (n^\pm)^{\nu_3} \\ \\
\qquad \qquad   k_1^\pm + k_2^\pm + k_3^\pm = 0  
   \end{array}
$ \\ 
\vspace{.3cm}
\parbox{1cm}{(d)}

  \caption{\small Feynman rules for the lowest-order effective vertices. Wavy lines denote reggeized fields and curly lines gluons. (a) is the direct transition vertex and (b) the reggeized gluon propagator. We also show the unregulated order $g$  (c) and order $g^2$ (d) induced vertices. }
  \label{fig:feynrules0p2}
\end{figure}

The lowest order Feynman rules for the induced vertices are shown in
{Fig}.~\ref{fig:feynrules0p2}. The contributions in the
vertices of the form $1/k^\pm$ generate a new type of divergencies which will be related to 
high energy logarithms. These divergencies call for a
regularization, or equivalently, a suitable definition of the nonlocal
operator $\partial_\pm^{-1}$ in the Wilson lines. A convenient
regularization scheme was defined in~\cite{Hentschinski:2011tz,Chachamis:2012gh} where $n^+$ and $n^-$ are replaced by tilted light-cone vectors of the form $a=n^-+e^{-\rho}n^+$ and $b=n^++e^{-\rho}n^-$. These tilted light-cone vectors form a hyperbolic angle $\rho$ in Minkowski space which can be interpreted as  $\ln s$, implying that at high energies we are interested in the $\rho\to\infty$ limit. In the following we treat  $\rho$  as  an external parameter which, at the end, we consider in the $\rho \to \infty$ limit, similar to the treatment of $\epsilon \to 0$ in $d=4 + 2 \epsilon$ dimensional regularization.  \\

After this brief Introduction we turn in the following to give details
of our calculation of the gluon-initiated forward jet
vertex. In Section~\ref{sec:virtual} we provide a description of the virtual
corrections to the gluon-gluon-reggeized gluon vertex, while in 
Section~\ref{sec:real} we explain the key details for the calculation of the
real corrections. Finally, Section~\ref{sec:concl} contains our conclusions
and an outlook for future calculations. The Appendix collects further technical details.

\section{Virtual Corrections to the Gluon-Gluon-Reggeized Gluon Vertex}
\label{sec:virtual}
\begin{figure}[htp]
\centering
\parbox{.35 \textwidth}{\includegraphics[width =.2 \textwidth]{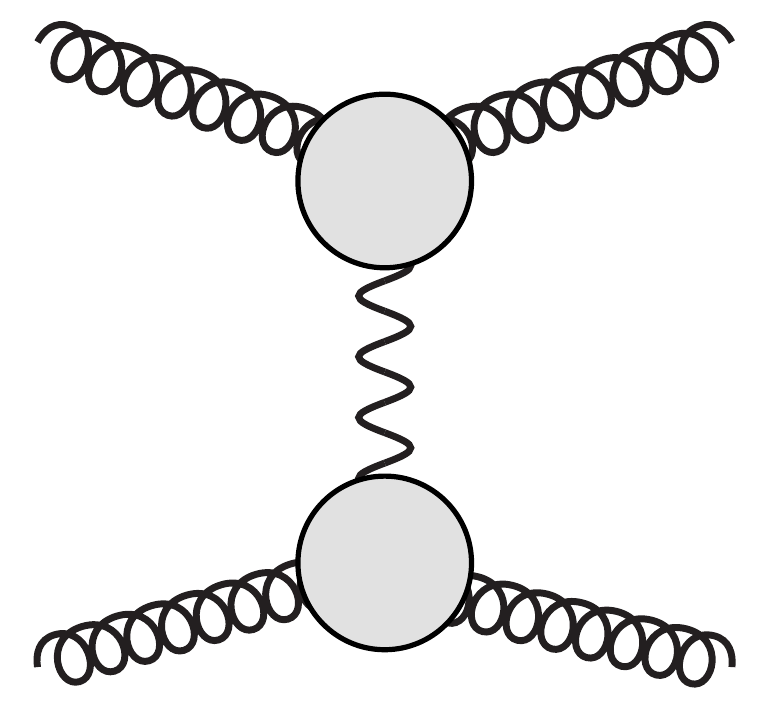}} \parbox{2cm}{
  \includegraphics[width =2cm]{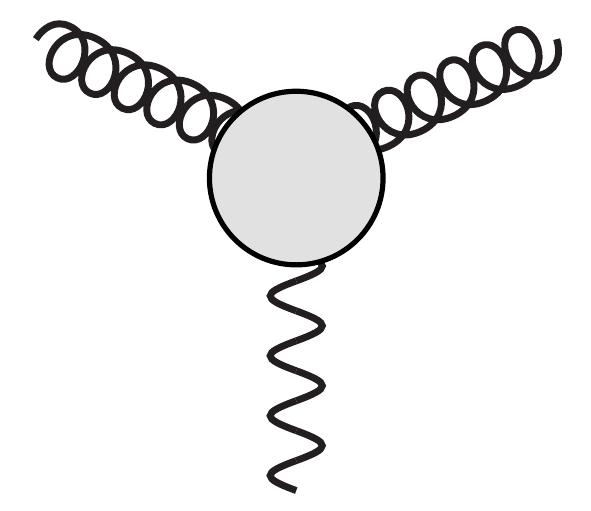} }$ = $
 \parbox{2cm}{
  \includegraphics[width =2cm]{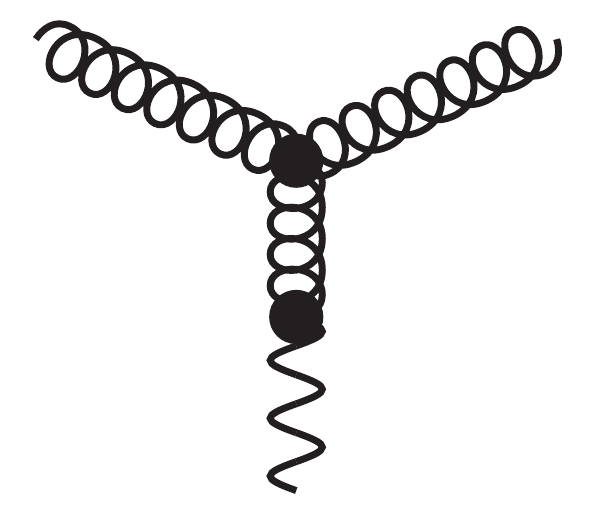} }$ +$
 \parbox{2cm}{
  \includegraphics[width =2cm]{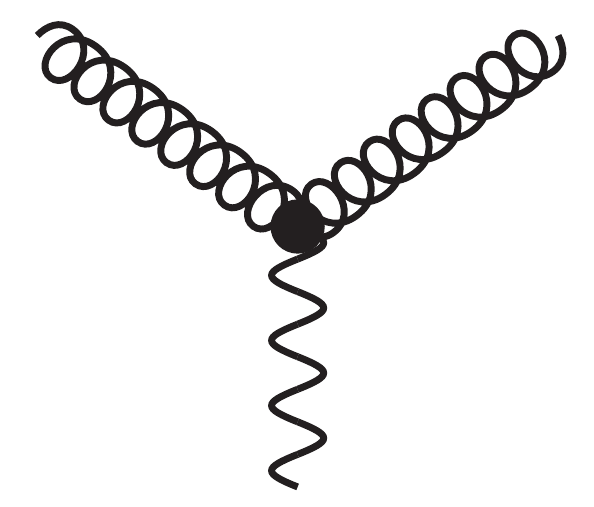} }
\caption{Tree-level contribution to the gluon-gluon scattering amplitude in terms of effective vertices.}
\label{tree}
\end{figure}

Let us consider the process $gg\to gg$, where the external gluons are on-shell: $p_a^2=(p_a-q)^2=p_b^2=(p_b+q)^2=0$. In the high-energy limit the corresponding scattering amplitude factorizes into a reggeized gluon  exchange in the $t$-channel and its couplings to the external particles, the so-called impact factors. At tree level this is shown in {Fig.}~\ref{tree}. \\
\\
Light cone momenta are defined making use of the momenta of the incoming
particles $p_a$ and $p_b$ through $p_a= p_a^+ n^- / 2$ and $p_b =
p_b^-n^+ / 2$ with $2 p_a \cdot p_b = s = p_a^+ p_b^-$. From
Eq.~(\ref{eq:constraint}) one obtains for the upper vertex the
constraint $q^+=0$, while $q^-=0$ for the lower one. Both constraints
can be understood as the leading term in the expansion in the small 
$q^+/p_a^+$ and $q^-/p_b^-$ ratios, respectively.  The polarization
vectors must be physical, satisfying for the upper vertex
$\varepsilon\cdot p_a=0$ and $\varepsilon^*\cdot(p_a-q)=0$. The last
relation implies that $\varepsilon^*\cdot p_a=\varepsilon^*\cdot q$.
Gauge invariance of the effective action enables us to choose {\it
  different} gauges for the upper and lower gluon-gluon-reggeized
gluon couplings. We therefore impose the condition
$\varepsilon(p_a)\cdot n^+=\varepsilon^*(p_1)\cdot n^+=0$ for the
upper vertex and the condition $\varepsilon(p_b)\cdot
n^-=\varepsilon^*(p_2)\cdot n^-=0$ for the lower vertex, which implies
the following polarization sum
\begin{equation}
\sum_{\text{polarizations}}\varepsilon_\mu^\lambda(p,n)(\varepsilon^{\lambda'}_\nu)^*(p,n) =  -g_{\mu\nu}+\frac{p_{\mu} n_\nu+p_{\nu} n_\mu}{p\cdot n},
\end{equation}
with $p$ being the gluon momentum and $n = n^\pm$.

To define the impact factors we start from the  general definition for 
 the differential cross-section for $m$-particle production in terms of the corresponding 
 matrix elements and the phase space integral, {\it i.e.}
\begin{equation}
d\sigma=\frac{1}{2s}|{\cal M}_{i\to f}|^2d\Pi_m (2\pi)^d\delta^d\bigg(p_a+p_b-\sum_{j=1}^m p_j\bigg);\quad d\Pi_m =\prod_{j=1}^m\frac{d^dp_j}{(2\pi)^{d-1}}
\delta_+ (p_j^2).
\end{equation}
In the special case where all final state particles are produced
either in the fragmentation region of particle $a$ or particle $b$ we
rewrite, with $m = m_a + m_b$, the overall delta function for global 
momentum conservation as follows
\begin{align}
  \label{eq:delta}
 (2\pi)^d\delta^d\bigg(p_a+p_b-\sum_{j=1}^m p_j\bigg)  &= \int\frac {d^d k}{(2 \pi)^d}  
  (2\pi)^{2d}\delta^{d}\bigg(p_a+k-\sum_{j=1}^{m_a} p_j\bigg)
\times
 \delta^d\left(p_b-k-\sum_{l=1}^{m_b}p_j\right).
\end{align}
The generalization to the additional production of $n$-particle clusters at central rapidities (see also  Fig.~\ref{fig:1}.a), with $m = m_a + m_b + \sum^n_i m_i$,  reads 
\begin{align}
  \label{eq:delta2}
& (2\pi)^d\delta^d\bigg(p_1+p_2-\sum_{j=1}^m p_j\bigg)  =\prod_{i = 0}^{n} \int\frac {d^d k_i}{(2 \pi)^d} 
  (2\pi)^{(2 + n)d}\delta^d\bigg(p_a+k_0-\sum_{l_0=1}^{m_a} p_{l_0}\bigg) \times
 \notag \\
&\delta^d \bigg(p_b-k_n-\sum_{l_n=1}^{m_b}p_j\bigg)
\times  (2\pi)^{n d}\delta^d\bigg(k_0-k_1-\sum_{l_1=1}^{m_1}p_{l_1}\bigg) \times \ldots 
 \delta^d\bigg(k_{n-1}-k_{n}-\sum_{l_{n}=1}^{m_{n}}p_{l_{n}}\bigg)
.
\end{align}
Restricting from now on to the $gg \to gg$ amplitude at tree-level, we
use in our next step the fact that the effective action naturally factorizes
the amplitude $ i{\cal M}_{gr^*\to g_1}$ into two products of $i{\cal
  M}_{gr^*\to g_1}$ times the square-root of the reggeized gluon
propagator $i/2{\bm q}^2$.  Squaring, averaging over color and
polarization of the initial gluon and summing over color and
polarization of the final state and reggeized gluon (at the level of
the $gr^* \to g$ amplitudes), the $2 \to 2$ tree-level amplitude 
takes the following factorized form
\begin{equation}
\overline{|{\cal M}^{(0)}_{g_ag_b\to g_1g_2}|^2}=\frac{\overline{|{\cal M}^{(0)}_{g_ar^*\to g_1}}|^2  }{2  \vec{k}^2 \sqrt{ N_c^2-1} } \times      \frac{\overline{|{\cal M}^{(0)}_{g_br^*\to g_2}}|^2    } {2  \vec{k}^2  \sqrt{ N_c^2-1}   }.
\end{equation}
Defining now the impact factors $=h_{a,b}^{(0)}(\vec{q})$ through the relation
\begin{equation}
{d\sigma_{ab}}=h_a^{(0)}(\vec{k})h_b^{(0)}(\vec{k})   {d^{2+2\epsilon}\vec{k}},
\end{equation}
we are lead to the following general expression in terms of the effective
action matrix elements:
\begin{align}
  \label{eq:impact_general}
  h_{a,{\rm gluon}}^{(0)}(\vec{k}) & = \frac{(2\pi)^{d/2}}{2 p_a^+}
  \int d k^- \frac{\overline{|{\cal M}^{(0)}_{g_ar^*\to g_1}}|^2 }{2
    \vec{k}^2 \sqrt{ N_c^2-1} } d \Pi_1 \delta^{(d)} (p_a + k - p_1 ),
\end{align}
with a natural 
generalization to $m_b$-particle production in the fragmentation region of particle $b$. 
 
\begin{figure}[h!]
\centering
\includegraphics[scale=.9]{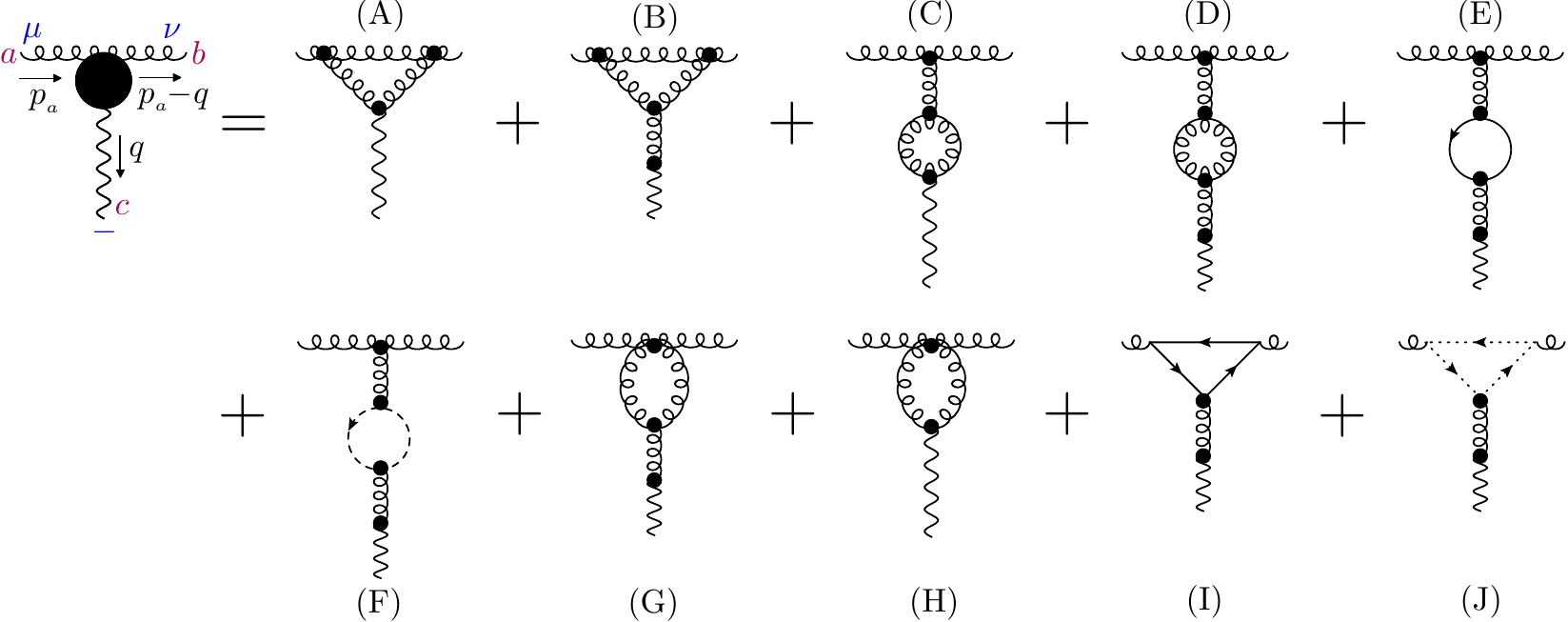}
\caption{One-loop virtual corrections to the gluon-initiated jet vertex.}
\label{fig:4}
\end{figure}

The $i{\cal M}_{gr^*\to g_1}$ amplitude itself receives at tree-level
two contributions (see {Fig.}~\ref{tree}): one from the
gluon-gluon-reggeized gluon (GGR) vertex 
$$g
  f_{abc}\frac{\vec{k}^2}{p_a^+}(n^+)^{\mu_1}(n^+)^{\mu_2}\varepsilon_{\mu_1}\varepsilon^*_{\mu_2}$$
and the other from the projection of the 3-gluon vertex 
$$g
  f_{abc}
  \left[2g^{\mu_1\mu_2}p_a^+-(n^+)^{\mu_1}(p_a-k)^{\mu_2}-(n^+)^{\mu_2}(2k+p_a)^{\mu_1}
  \right] \varepsilon_{\mu_1}\varepsilon^*_{\mu_2}. $$  
  It is possible to verify that this amplitude is gauge invariant and
satisfies the necessary Ward/Slavnov-Taylor identities, see {\it e.g.}
\cite{Lipatov:1995pn,Chachamis:2012mw}.  At  the amplitude level we arrive at
\begin{align}
  \label{eq:Mrgg}
  \mathcal{M}_{g_ar^*\to g_1} & = 2 g f_{abc} \epsilon\cdot\epsilon^*.
\end{align}
and
\begin{equation}\label{ggr}
h_{a,{\rm gluon}}^{(0)}(\vec{k})=\frac{N_c}{\sqrt{N_c^2-1}}\frac{g^2}{\vec{k}^2}\frac{1}{(2\pi)^{1+\epsilon}}=\frac{2^{1+\epsilon}\alpha_s C_A}{\mu^{2\epsilon}\Gamma(1-\epsilon)\sqrt{N_c^2-1}}\frac{1}{\vec{k}^2};\qquad\alpha_s\equiv\frac{g^2\mu^{2\epsilon}\Gamma(1-\epsilon)}{(4\pi)^{1+\epsilon}}.
\end{equation}
The one-loop corrections to this gluon-gluon-reggeized gluon vertex are shown in Fig.~\ref{fig:4}. All diagrams are evaluated in the limit $\rho \to \infty$, while we only keep track of
divergent ($\mathcal{O}(\rho)$) and finite ($\mathcal{O}(\rho^0)$)
terms; $\epsilon$ is on the other hand kept finite. Details about the
calculation of individual diagrams can be found in the appendix
\ref{appvirt}.  The final result for the 1-loop $gr^* \to g$ amplitude
reads
\begin{align}
i &{\cal M}^{(1)}_{g_ar^*\to g_1} =-\frac{g^3 \mu^{2 \epsilon} p_a^+}{(4\pi)^{2+\epsilon} }f_{abc} \left( \frac{{\vec k}^2}{\mu^2}\right)^\epsilon
\Bigg\{N_c\,\varepsilon\cdot\varepsilon^*\frac{\Gamma(1-\epsilon)\Gamma^2(\epsilon)}{\Gamma(2\epsilon)}
\bigg[2\ln\left(\frac{p_a^+}{|\vec{k}|}\right)+\rho+\psi(1)
\notag \\ & 
-2\psi(\epsilon)+\psi(1-\epsilon)\bigg] +8 [N_c(1+\epsilon)-n_f]\bigg[-\frac{1}{2}\varepsilon\cdot\varepsilon^*\frac{\Gamma^2(1+\epsilon)\Gamma(-\epsilon)}{\Gamma(4+2\epsilon)}
\notag \\
&
+\frac{\varepsilon\cdot q\,\varepsilon^*\cdot q}{\vec{k}^2}\frac{\Gamma(\epsilon)\Gamma(1-\epsilon)}{\Gamma(4+2\epsilon)}(2\Gamma(1+\epsilon)+\Gamma(2+\epsilon))\bigg] 
\notag \\
&
+8[N_c(1+\epsilon)-n_f]\frac{\varepsilon\cdot q\,\varepsilon^*\cdot q}{\vec{k}^2}\frac{\Gamma(-\epsilon)\Gamma^2(1+\epsilon)}{(2+2\epsilon)\Gamma(2+2\epsilon)}+\varepsilon\cdot\varepsilon^*(4N_c-n_f)\frac{\Gamma(-\epsilon)\Gamma^2(1+\epsilon)}{\epsilon\,\Gamma(2+2\epsilon)}\nonumber\\
&-\varepsilon\cdot\varepsilon^*(4N_c-n_f)\frac{1+2\epsilon}{\epsilon}\frac{\Gamma(-\epsilon)\Gamma^2(1+\epsilon)}{\Gamma(2+2\epsilon)}+2N_c\,\varepsilon\cdot\varepsilon^*\frac{\Gamma(-\epsilon)\Gamma^2(1+\epsilon)}{\Gamma(2+2\epsilon)}\nonumber\\
&+2\,\varepsilon\cdot\varepsilon^*[(1+\epsilon)N_c-n_f]\frac{\Gamma(-\epsilon)\Gamma^2(1+\epsilon)}{(3+2\epsilon)\Gamma(2+2\epsilon)}-2\varepsilon\cdot\varepsilon^*[4N_c-n_f]\frac{\Gamma(-\epsilon)\Gamma^2(1+\epsilon)}{\Gamma(2+2\epsilon)}\Bigg\}. 
\end{align}
Our result contains  two types of tensor structures involving the initial and final polarization tensors, namely,  $\varepsilon\cdot\varepsilon^*$ and 
$\varepsilon\cdot q\,\varepsilon^*\cdot q / \vec{k}^2$. These are related to the helicity conserving and helicity violating terms~\cite{Fadin:1999de}
\begin{equation}
\label{eq:deltahel}
\varepsilon\cdot\varepsilon^*= \delta_{\lambda_a,\lambda_{1}};\qquad \varepsilon\cdot\varepsilon^*+ \frac{2}{\vec{k}^2}\varepsilon\cdot q\,\varepsilon^*\cdot q =- \delta_{\lambda_a,-\lambda_{1}}.
\end{equation}
To avoid double counting it is now needed to introduce the subtraction
procedure discussed in the Introduction.  This procedure requires to
subtract all effective diagrams which contain internal reggeized gluon
propagators in the $t$-channel from the above result.  With
\begin{align}
  \label{eq:imp2}
  \parbox{2cm}{\includegraphics[width=2cm]{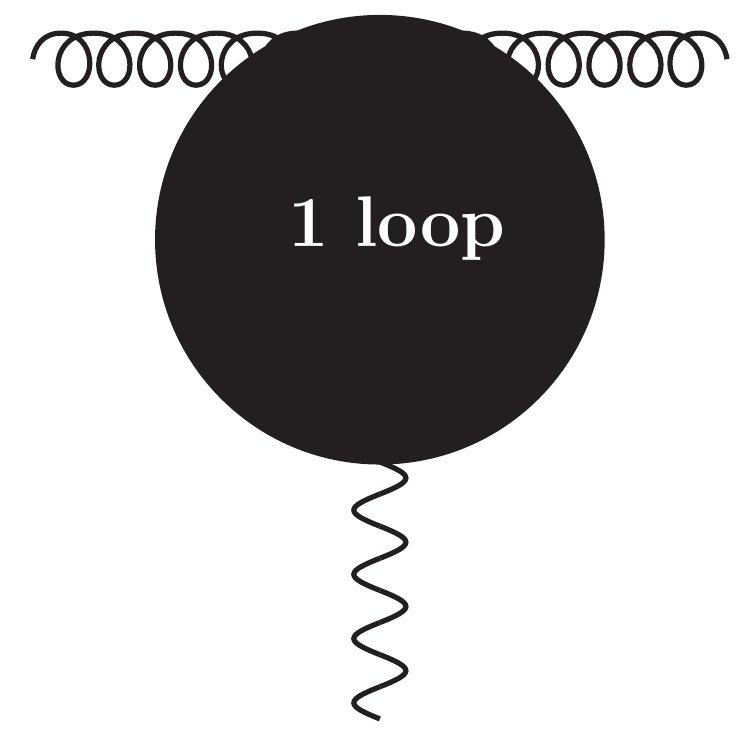}} = i{\cal M}^{(1)}_{g_ar^*\to g_1},
\end{align}
the subtraction procedure results into  the following coefficient
\begin{equation}
 \mathcal{C}_{gr^* \to g} \left(  \frac{ p_a^+}{\sqrt{{\bm k}^2}}, \rho; \epsilon \frac{{\bm k}^2}{\mu^2} \right)  = \parbox{2cm}{\includegraphics[width=2cm]{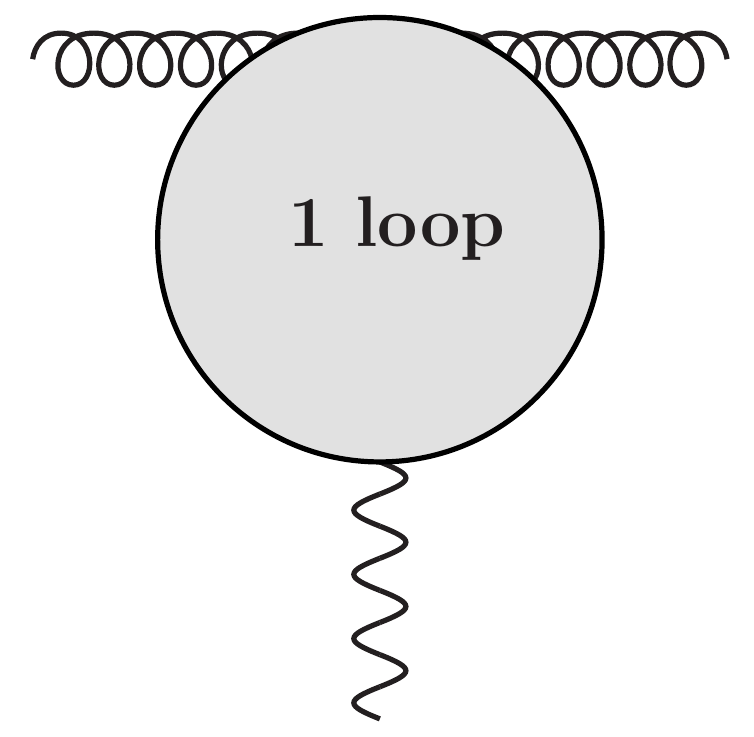}}  = \parbox{2cm}{\includegraphics[width=2cm]{impaamp1l.pdf}}  - \parbox{2cm}{\includegraphics[width=2cm]{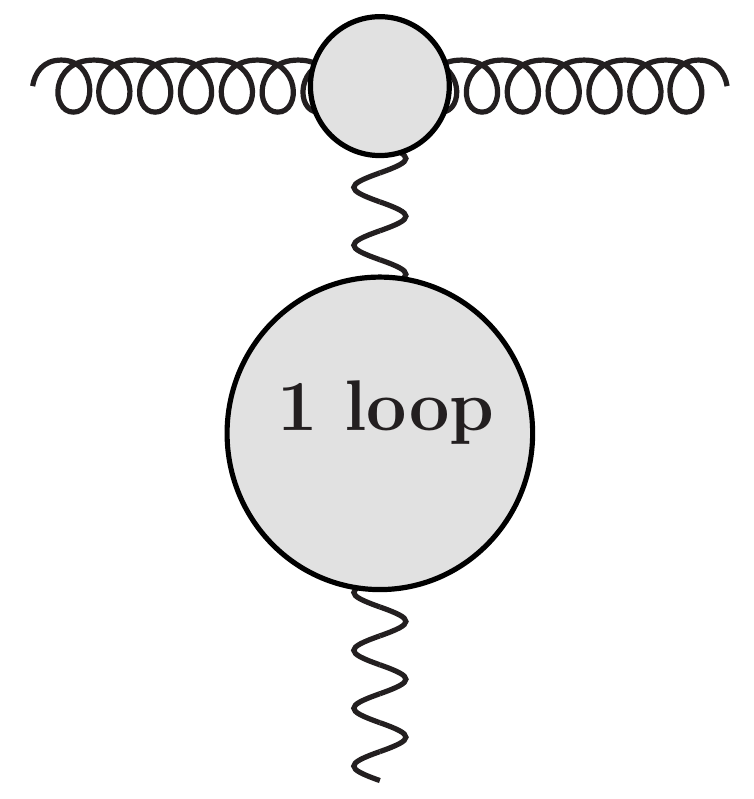}} .
\end{equation}
The one-loop reggeized gluon self-energy, which was computed in~\cite{Hentschinski:2011tz}, includes both divergent, $\mathcal{O}(\rho)$, and finite terms of ${\cal O}(\rho^0)$ and reads\footnote{There is a misprint in~\cite{Hentschinski:2011tz}: the term $[\cdots]_2$ in Eq. 6 of \cite{Hentschinski:2011tz} vanishes. As the same contribution appears also in the 1-loop corrections to the quark-quark-reggeized gluon vertex, the final result is independent of this contribution and remains unchanged.}
\begin{equation}
\Sigma^{(1)}(\rho; \epsilon, \frac{{\vec k}^2}{\mu^2}) = \vcenter{\hbox{\includegraphics[scale=0.25]{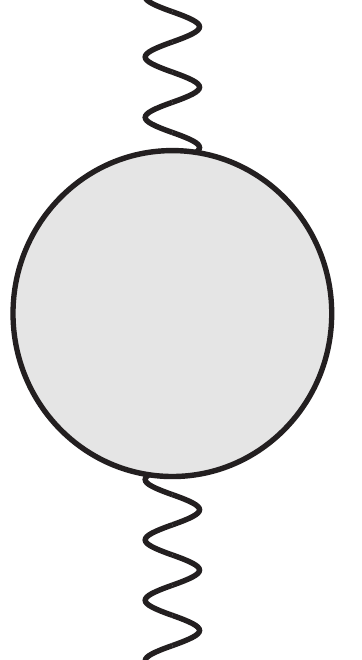}}}=(-2i\vec{k}^2) \, \omega^{(1)}(\vec{k}^2)\left[\rho-\frac{i\pi}{2}+\frac{ 5+3\epsilon-\frac{n_f}{N_c}(2+2\epsilon)}{2(1+2\epsilon)(3+2\epsilon)}\right]. 
\end{equation}
where  
\begin{align}
  \label{eq:omega1}
  \omega^{(1)}(\vec{k}^2)  & =-\frac{\alpha_s N_c}{2\pi}\frac{\Gamma^2( 1 + \epsilon)}{\epsilon \Gamma(1 + 2\epsilon)}\left(\frac{\vec{k}^2}{\mu^2}\right)^\epsilon
\end{align}
  is the 
one-loop gluon Regge trajectory.
The  high energy limit of the   gluon-gluon scattering amplitude 
at one-loop  is then obtained as the following sum of diagrams
\begin{equation}\label{trajectory}
\parbox{3cm}{\center \includegraphics[width = 2cm]{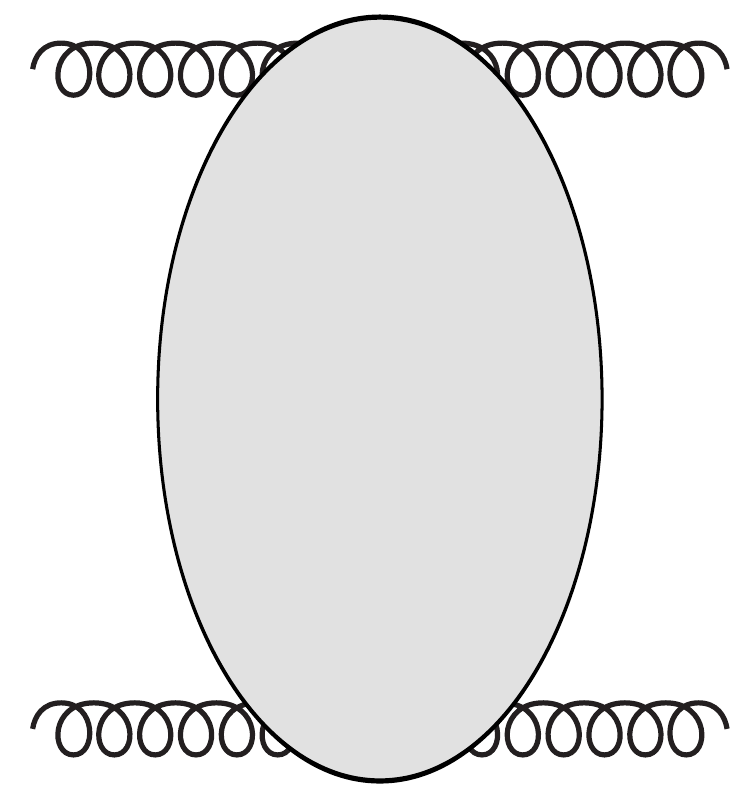}}
 = 
\parbox{3cm}{\center \includegraphics[width = 2cm]{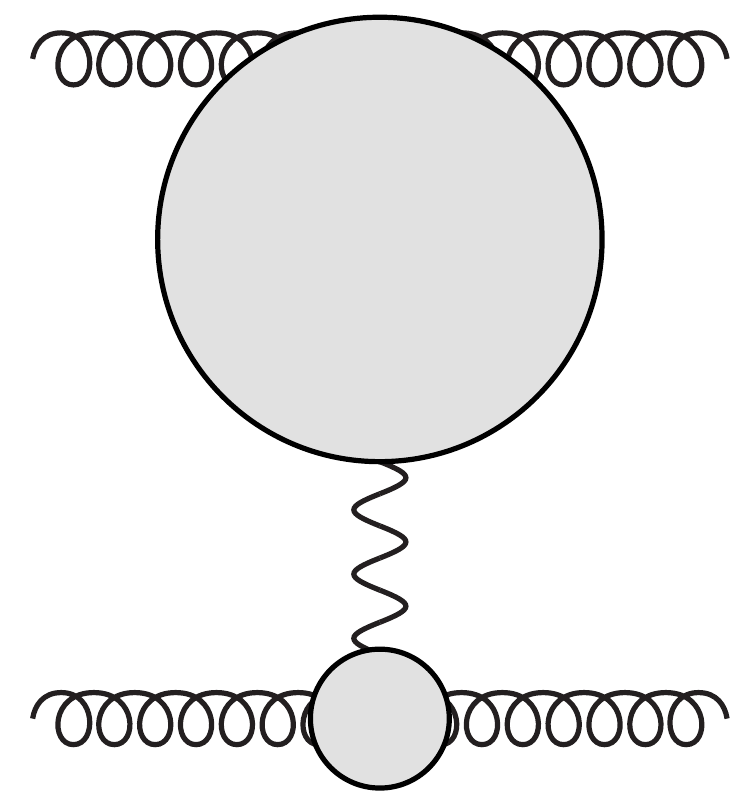}}
+
\parbox{3cm}{\center \includegraphics[width = 2cm]{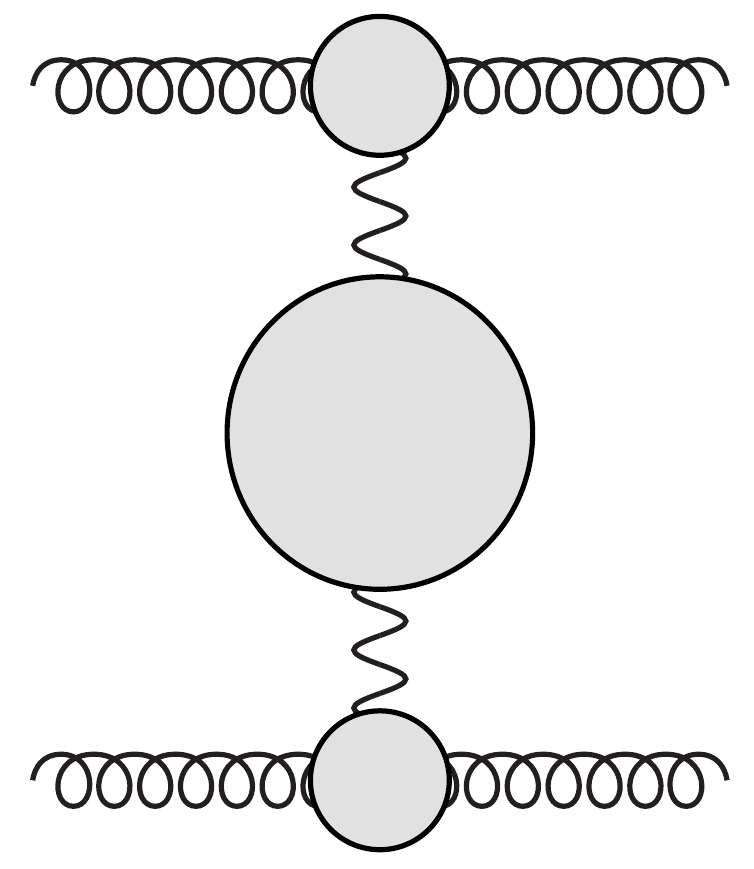}}
+
\parbox{3cm}{\center \includegraphics[width = 2cm]{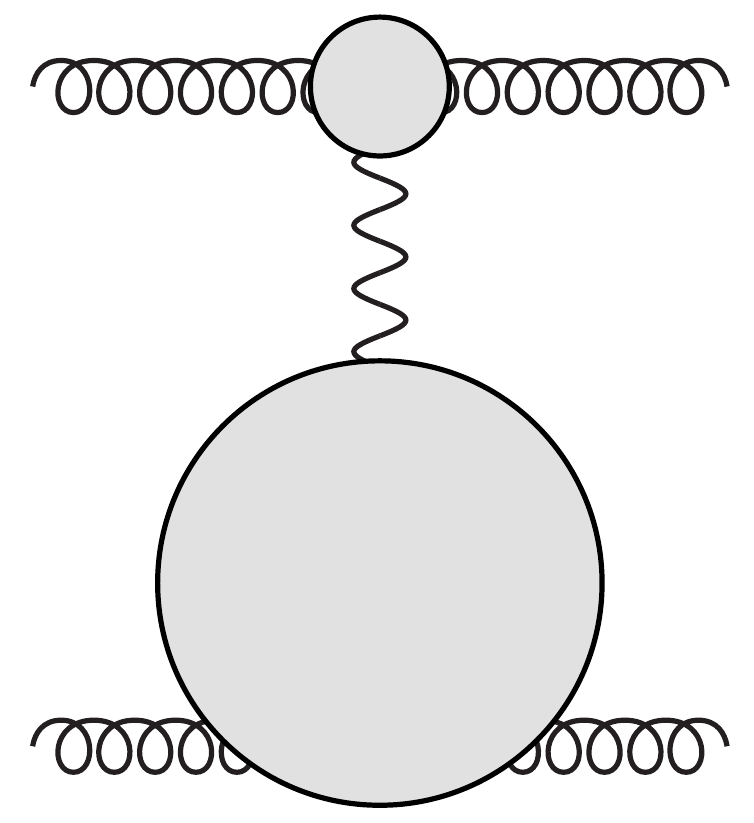}}
\end{equation}
While each diagram on the right side is divergent in the limit $\rho
\to \infty$, the divergence cancels in their sum, resulting into a
finite one-loop amplitude. Following~\cite{Chachamis:2012gh} we
therefore define renormalized gluon-gluon-reggeized gluon coupling
coefficients,
\begin{align}
  \label{eq:renomcoeff}
   \mathcal{C}^{ \text{R}}_{gr^* \to g} \left( \frac{ p_a^+}{M^+}; \epsilon,  \frac{{\bm q}^2}{\mu^2} \right) 
& =
 Z^+ \left( \frac{M^+}{\sqrt{{\bm k}^2}}, \rho ; \epsilon,  \frac{{\bm k}^2}{\mu^2}  \right)   \mathcal{C}_{gr^* \to g} \left(  \frac{ p_a^+}{\sqrt{{\bm k}^2}}, \rho; \epsilon \frac{{\bm k}^2}{\mu^2} \right), 
\\
   \mathcal{C}^{ \text{R}}_{gr^* \to g} \left( \frac{ p_b^-}{M^-}; \epsilon,  \frac{{\bm k}^2}{\mu^2} \right) 
& =
 Z^- \left( \frac{M^-}{\sqrt{{\bm k}^2}}, \rho ; \epsilon,  \frac{{\bm k}^2}{\mu^2}  \right)   \mathcal{C}_{gr^* \to g} \left(  \frac{ p_b^-}{\sqrt{{\bm k}^2}}, \rho; \epsilon \frac{{\bm k}^2}{\mu^2} \right), 
\end{align}
and the renormalized reggeized gluon propagator,
\begin{align}
  \label{eq:renompropR}
  G^{\text{R}} \left(M^+, M^-; \epsilon, {\bm k}^2, \mu^2 \right)  & = 
\frac{ G \left(\rho; \epsilon, {\bm k}^2, \mu^2   \right) }{ Z^+ \left( \frac{M^+}{\sqrt{{\bm k}^2}}, \rho ; \epsilon,  \frac{{\bm k}^2}{\mu^2}  \right) Z^- \left( \frac{M^-}{\sqrt{{\bm k}^2}}, \rho ; \epsilon,  \frac{{\bm k}^2}{\mu^2}  \right) },
\end{align}
with the bare reggeized gluon propagator given by
\begin{align}
  \label{eq:barepropR}
   G \left(\rho; \epsilon, {\bm k}^2, \mu^2   \right)
&=
\frac{i/2}{{\bm k}^2} \left\{ 1 + \frac{i/2}{{\bm k}^2}
 \Sigma \left(\rho; \epsilon, \frac{{\bm k}^2}{\mu^2}    \right)  + \left[  \frac{i/2}{{\bm k}^2} \Sigma \left(\rho; \epsilon, \frac{{\bm k}^2}{\mu^2}   \right)\right] ^2 + \ldots   \right\}. 
\end{align}
The renormalization factors $Z^\pm$ cancel for the complete  scattering amplitude and can be  parametrized as follows
\begin{align}
  \label{eq:Z+-}
  Z^\pm\left( \frac{M^\pm}{\sqrt{{\bm k}^2}}, \rho ; \epsilon,  \frac{{\bm k}^2}{\mu^2}  \right)  & = \exp \left [ \left(\rho -  \ln \frac{M^\pm}{\sqrt{\bm k}^2}  \right) \omega\left(\epsilon, \frac{{\bm k}^2}{\mu^2} \right) + f^\pm \left(\epsilon, \frac{{\bm k}^2}{\mu^2} \right) \right],
\end{align}
where the gluon Regge trajectory has the following perturbative expansion,
\begin{align}
  \label{eq:omega_expand}
   \omega\left(\epsilon, \frac{{\bm k}^2}{\mu^2} \right) & =  \omega^{(1)}\left(\epsilon, \frac{{\bm k}^2}{\mu^2} \right) + \omega^{(2)}\left(\epsilon, \frac{{\bm k}^2}{\mu^2} \right) + \ldots,
\end{align}
with the one-loop expression given in Eq.~\eqref{eq:omega1}.  The
function $f(\epsilon, {\bm k}^2)$ parametrizes finite contributions
and is, in principle, arbitrary.  While symmetry of the scattering
amplitude requires $f^+ = f^- = f$, Regge theory suggests to fix it in
such a way that at one loop the non-$\rho$-enhanced contributions of
the one-loop reggeized gluon self energy are entirely transferred to
the quark-reggeized gluon couplings. This leads to
\begin{align}
  \label{eq:f1loop}
   f^{(1)}\left(\epsilon, \frac{{\bm k}^2}{\mu^2} \right) & =  -\frac{ \alpha_s N_c \Gamma^2(1 + \epsilon)}{4 \pi \Gamma(1 + 2 \epsilon)} \left(\frac{{\bm q}^2}{\mu^2} \right)^\epsilon  
        \bigg[  \frac{5 + 3\epsilon  -\frac{n_f}{N_c} (2 + 2\epsilon)}{2(1 + 2 \epsilon)(3 + 2 \epsilon)} 
\bigg]  .
\end{align}
Let us remark that in principle other alternative and even asymmetric  $f^+ \neq f^-$  choices are possible, as long as they are in agreement with UV-renormalizability of QCD and collinear factorization.  
Using now  $M^+ = p_a^+$, $M^- = p_b^-$ we can see that this choice for $f$ keeps the full $s$-dependence of 
the amplitude  inside the reggeized gluon exchange.  The renormalized
gluon-gluon-reggeized gluon couplings  allows then to extract the NLO corrections to the gluon impact factor. Extracting the Born contribution and decomposing into helicity conserving and non-conserving parts
\begin{align}
  \label{eq:rnoG}
  \mathcal{C}^{ \text{R}}_{gr^* \to g} \left( 1; \epsilon,  \frac{{\bm q}^2}{\mu^2} \right)  & =  2 g f_{abc} \cdot  \left[  \Gamma_a^{(+)}  \delta_{\lambda_a,\lambda_1} +  \Gamma_a^{(-)} \delta_{\lambda_a,-\lambda_1}   \right],
\end{align}
 where the helicity tensors are for finite $\epsilon$ defined through Eq.~(\ref{eq:deltahel}), we have 
\begin{equation}
\begin{aligned}
\label{eq:virtresu}
\Gamma_a^{(+)}&=-\frac{1}{2}\omega^{(1)}\left[-\psi(1)+2\psi(2\epsilon)-\psi(1-\epsilon)+\frac{1}{4(1+2\epsilon)(3+2\epsilon)}+\frac{7}{4(1+2\epsilon)}-\frac{n_f}{N_c}\frac{1+\epsilon}{(1+2\epsilon)(3+2\epsilon)}\right]
 \\
&= \frac{\alpha_s N_c}{4 \pi} \left(\frac{{\bm q}^2}{\mu^2} \right)^\epsilon
 \left[
- \frac{1}{\epsilon^2} + \frac{\beta_0}{2 \epsilon} - \frac{(67 - \pi^2) N_c - 10 n_f }{18}
\right] + \mathcal{O}(\epsilon)
,  \qquad \qquad \beta_0 = \frac{11}{3}N_c - \frac{2}{3}n_f;\\
\Gamma_a^{(-)}&=-\frac{1}{2}\omega^{(1)}\left[\frac{\epsilon}{(1+\epsilon)(1+2\epsilon)(3+2\epsilon)}\left(1+\epsilon-\frac{n_f}{N_c}\right)\right] 
=
\frac{\alpha_s}{12 \pi} (N_c - n_f) + \mathcal{O}(\epsilon).
\end{aligned}
\end{equation}
which is in  precise agreement with the literature~\footnote{As noted in~\cite{DelDuca:1998kx}, the original result~\cite{Fadin:1993wh} contains several misprints. The correct expressions can be found {\it e.g.} in~\cite{Lipatov:1996ts, DelDuca:1998kx, Fadin:1999de}. }~\cite{Fadin:1993wh,Fadin:1992zt, Lipatov:1996ts, DelDuca:1998kx, Fadin:1999de}.

\section{Real Corrections and One-Loop Jet Vertex}
\label{sec:real}
\begin{figure}[htp]
\centering
\parbox{3cm}{\center \includegraphics[width = 2cm]{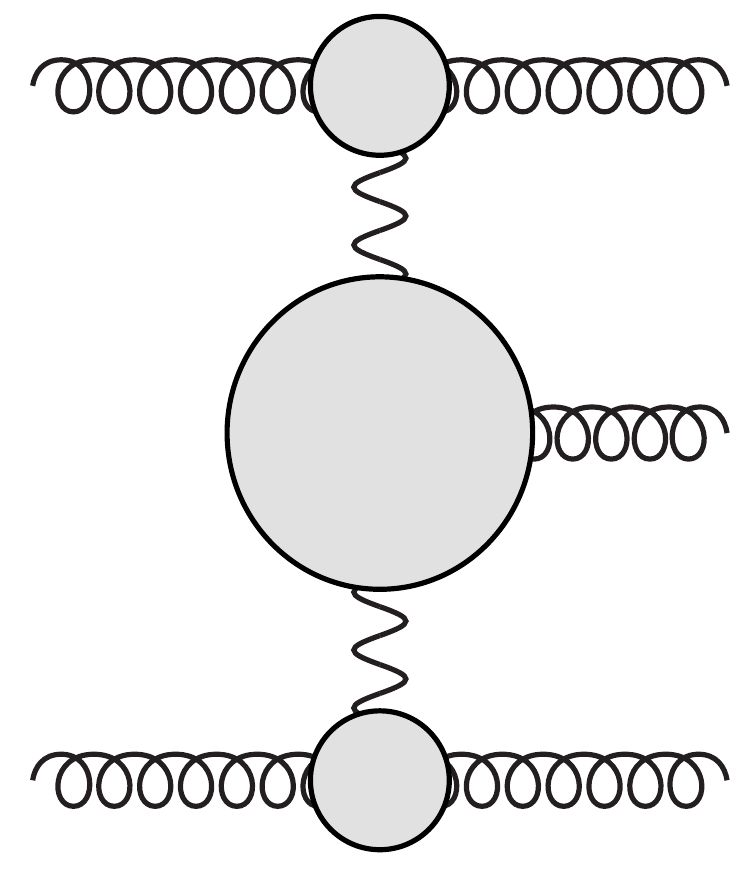}}
+
\parbox{3cm}{\center \includegraphics[width = 2cm]{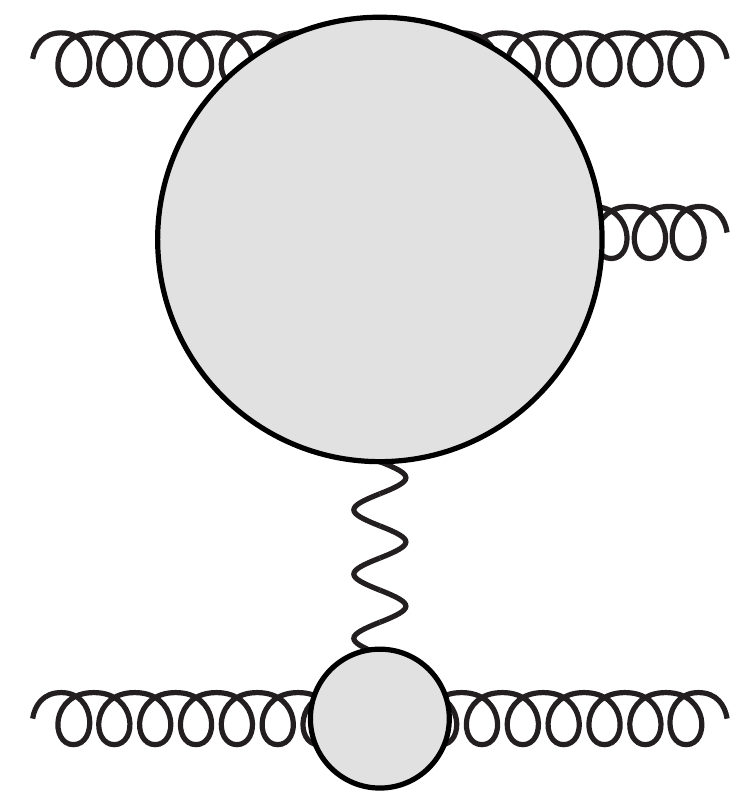}}
+
\parbox{3cm}{\center \includegraphics[width = 2cm]{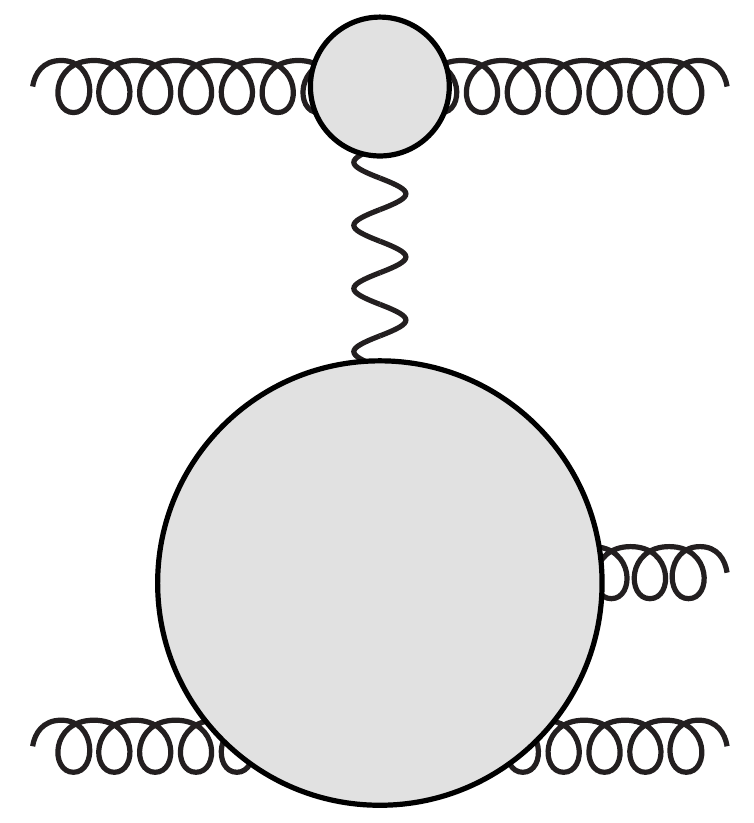}}
\caption{Central (left) and quasi-elastic (middle and right) gluon production.}
\label{fig:5}
\end{figure}

\begin{figure}[htp]
\centering
\parbox{3cm}{\center \includegraphics[width = 2cm]{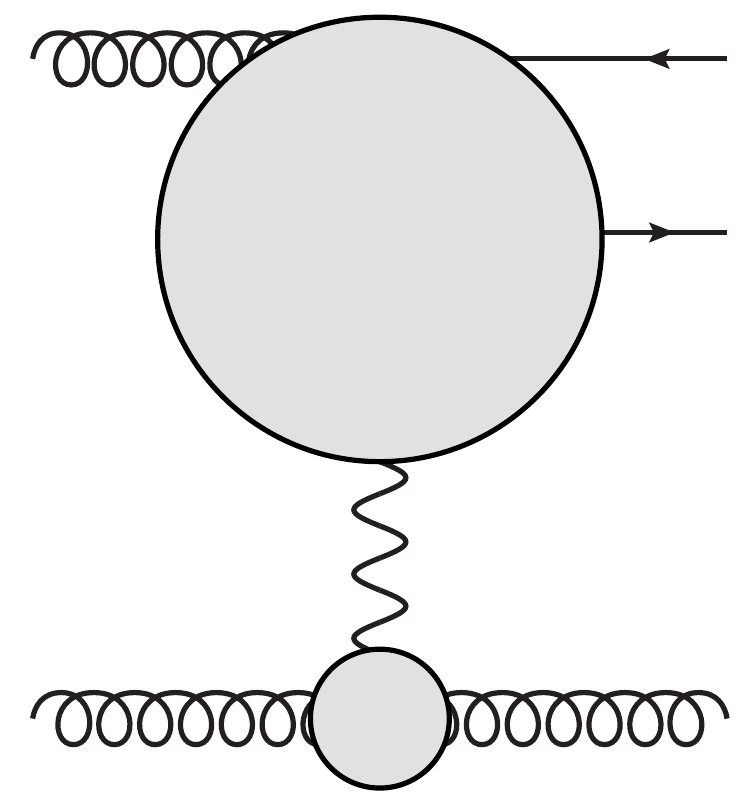}}
+
\parbox{3cm}{\center \includegraphics[width = 2cm]{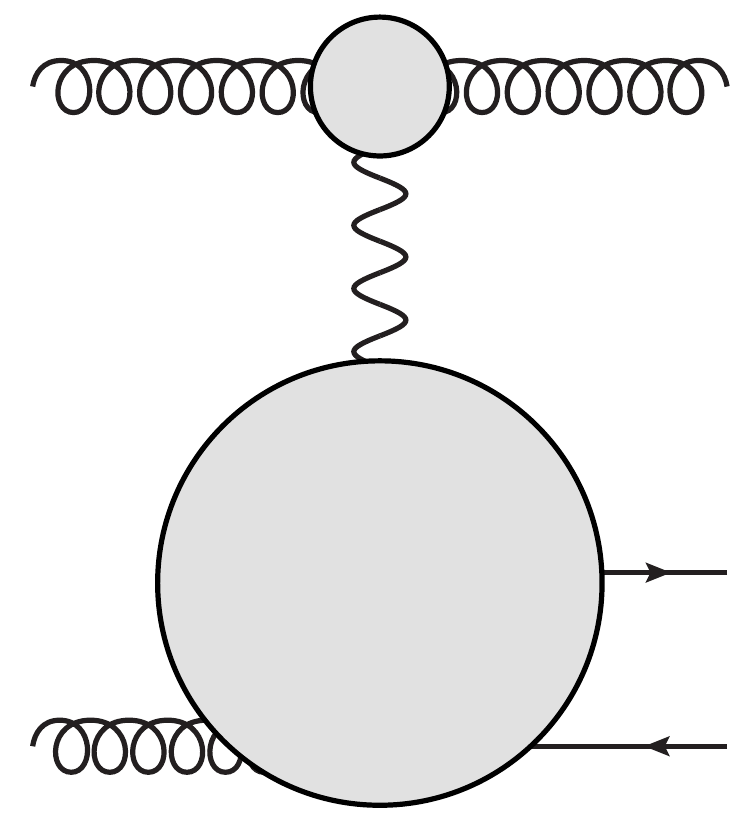}}
\caption{Quark production is restricted to the  quasi-elastic region.}
\label{fig:5b}
\end{figure}

In the high energy limit, the  real corrections to the Born-level process are naturally cast  into
three contributions to the $gg \to ggg$  amplitude where the additional gluon is either produced at  central rapidities or close to the fragmentation region of one of the initial gluons (quasi-elastic gluon production), see {Fig.}~\ref{fig:5}. A second class of corrections is due to the possible fragmentation of one of the initial gluons into a $q\bar{q}$ pair which only contributes to the quasi-elastic region, see  {Fig.}~\ref{fig:5b}.
In the same way as the effective action generates high energy
divergences near the light-cone when computing virtual corrections, a
cut-off in
rapidity must be enforced in the longitudinal integrations.\\

The central production amplitude yields the unintegrated
real part of the forward leading order BFKL kernel
and is obtained from the sum of the following three effective
diagrams:
\begin{equation}\label{qe}
\vcenter{\hbox{\includegraphics[scale=0.3]{central2}}}.
\end{equation}
The squared amplitude for \eqref{qe}, averaged over color of the
incoming reggeized fields and summed over final state color and
helicities reads~\cite{Hentschinski:2011tz}
\begin{equation}
\overline{|{\cal M}|^2}_{r^*r^*\to g}=\frac{16 g^2N_c}{N_c^2-1}\frac{{\vec{k}'}^2\vec{k}^2}{\vec{q}^2}
\end{equation}
with ${{\vec k}' = {\vec k} - {\vec q}}$. 
Defining the central production vertex  $V^{(0)}$ through the   differential cross section\footnote{Strictly speaking we cannot identify the gluon with transverse momentum ${\vec q} = {\vec k} - {\vec k'}$ to be the centrally produced one. The cross-section should be therefore written as the sum over the three possibilities where one of the gluons is produced at central rapidities, together with the corresponding symmetry factor $1/3$.},
\begin{equation}\label{eq:decomV}
d\hat{\sigma}_{ab}^{(c)}=h_a^{(0)}(\vec{k}')h_b^{(0)}(\vec{k}){\cal V}(\vec{k},\vec{k}'; {\vec q})d^{2+2\epsilon}\vec{k}'\,d^{2+2\epsilon}\vec{k}\,dy,
\end{equation}
we obtain, using the decomposition in Eq.~(\ref{eq:delta2}) and with $\mu^2 = {k^+}' k^-$:
\begin{align}
  \label{eq:defV}
  V(\vec{k},\vec{k}'; {\vec q}) &= \frac{N_c^2 -1}{8 } \int {d\mu^2}   \frac{\overline{|\mathcal{M}|^2}_{r^*r^* \to g}}{ {\vec k}^2 {\vec k'}^2} d \Pi_1  \delta^{(d)}(k' - k - q) 
\notag \\
&
=
\frac{\alpha_sN_c}{\pi_\epsilon \pi{\vec{q}^2}};\quad \qquad \qquad  \qquad  \pi_\epsilon\equiv \pi^{1+\epsilon}\Gamma(1-\epsilon)\mu^{2\epsilon}.
\end{align}
While the rapidities of the gluons in the forward and backward
directions are determined by kinematics, $y_a = \ln p_a^+/ |{\vec k'}|$
and $y_b = -\ln p_b^-/|{\vec{k}}|$, the integral over the gluon
rapidity in Eq.~(\ref{eq:decomV}) leads to a divergence in the limit $y \to \pm
\infty$ of inclusive observables. We therefore
introduce upper and lower bounds $\eta_{a,b}$ on this integral with
$\eta_a > y > \eta_b$ which we evaluate in the limit $\eta_{a,b} \to
\pm \infty$. This leads then to the definition of the regularized
production vertex ${\cal V}(\vec{k},\vec{k}'; {\vec q};
\eta_a,\eta_b)\equiv
V(\vec{k},\vec{k}')\Theta(\eta_a-y)\Theta(y-\eta_b)$.

\begin{figure}[htp]
\centering
\includegraphics[scale=0.5]{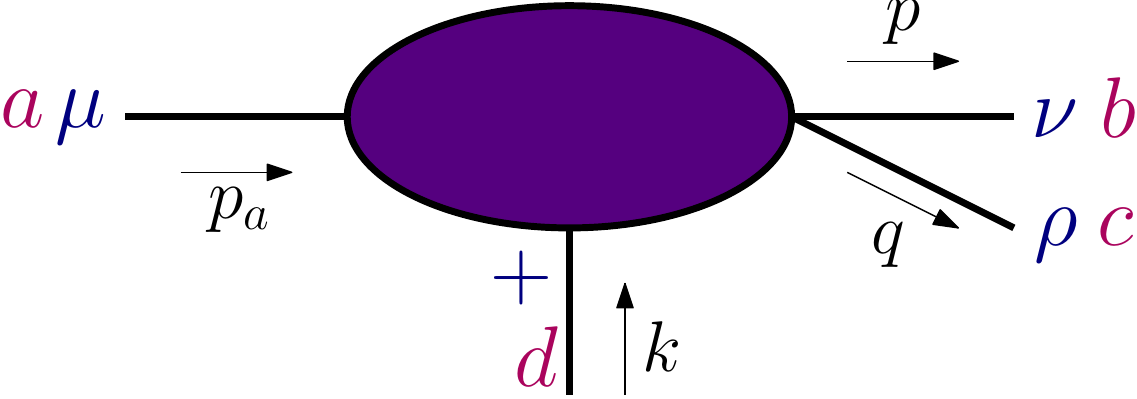}
\caption{Notation for external momenta and color indices in the quasi-elastic contribution.}
\label{fig:6}
\end{figure}

\begin{figure}[htp]
\centering
\includegraphics[scale=.8]{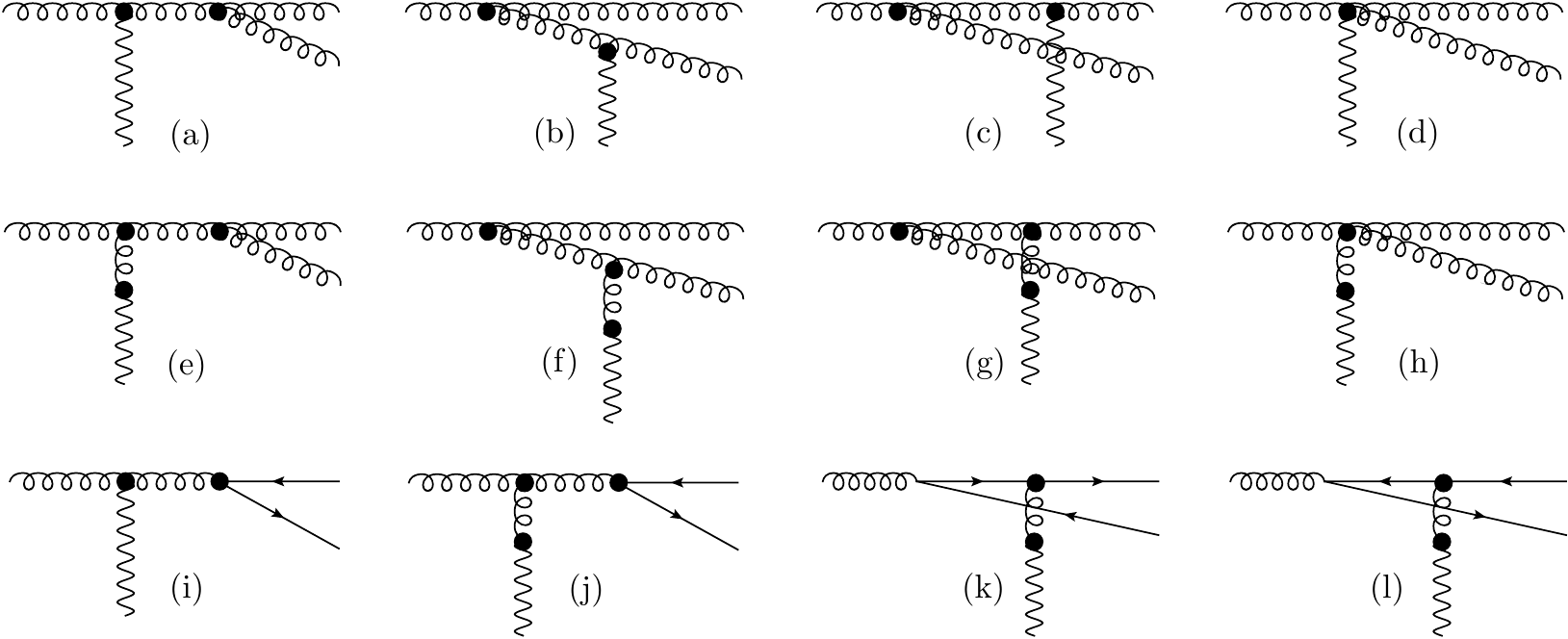}
\caption{Diagrams involved in the computation of the real corrections. In the case of final $q\bar{q}$ state, the external quark has momentum $p$ and the external antiquark momentum $q$.}
\label{fig:7}
\end{figure}

The remaining part of our calculation, the quasi-elastic contribution $g(p_a)r^*(k)\to g(p)g(q)$ (see the notation in {Fig.}~\ref{fig:6}), is given by the sum of all the effective diagrams 
in~{Fig.}~\ref{fig:7}, where we include both the $gg$ and the $q\bar{q}$ final state. 
In the computation of these diagrams we will employ the following choice for the polarization vectors: $\varepsilon_a\cdot p_a=\varepsilon_b\cdot p=\varepsilon_c\cdot q=\varepsilon_a\cdot n^+=\varepsilon_b\cdot n^+=\varepsilon_c\cdot n^+=0$; together with the Sudakov decomposition
\begin{equation}
\begin{aligned}
p_a&=p_a^+\frac{n^-}{2},\quad \quad p= (1-z) p_a^+\frac{n^-}{2}+p^-\frac{n^+}{2}+  {\vec k}',\\
k&=k^-\frac{n^+}{2}+ { \vec k},\quad q= z p_a^+ \frac{n^-}{2}+(k^--p^-)\frac{n^+}{2}+ {\vec q},
\end{aligned}
\end{equation}
where we used $q^+=z p_a^+$ and $p_a^+=(1-z)p_a^+$. Squaring and averaging over initial colors and polarizations and summing over the final state we obtain with ${\vec \Delta} = {\vec q} - z {\vec k}$  
\begin{equation}
\begin{aligned}
&\overline{|{\cal M}^2|}_{gg\to ggg}=
 \\ & \qquad 
8z(1-z) (p_a^+)^2
{g^4C_a}
(1+\epsilon)
{\cal P}_{gg}(z)\frac{ {\vec k}^2}{\vec{k'}^2  } \left[\frac{z^2\vec{k'}^2+(1-z)^2\vec{q}^2-z(1-z)\vec{q}\cdot \vec{k'} )}{\vec{q}^2\vec{\Delta}^2}\right],
\end{aligned}
\end{equation}
where ${\cal P}_{gg}(z)=C_a\frac{1+z^4+(1-z)^4}{z(1-z)}$ is the
gluon-gluon Altarelli-Parisi splitting function. In the same way, the
joined contribution of the diagrams (j), (k) and (l) for the
quark-antiquark final state gives 
\begin{equation}
\begin{aligned}
\overline{|{\cal M}^2|}_{gg\to q\bar{q}}&= 8z(1-z)  (p_a^+)^2  g^4  n_fN_c(1+\epsilon){\cal P}_{qg}(z,\epsilon)
 \frac{{\vec k}^2}{{\vec q}^2 {\vec k'}^2 }
\left[\frac{C_f}{C_a}+z(1-z)\frac{\vec{q}\cdot \vec{\Delta}}{ \vec{\Delta}^2}\right],
\end{aligned}
\end{equation}
with ${\cal P}_{qg}(z,\epsilon)=\frac{1}{2}\left[1-\frac{2z(1-z)}{1+\epsilon}\right]$  the quark-gluon splitting function. Generalizing our definition in Eq.~(\ref{eq:impact_general}) to two final state particles we obtain
\begin{align}
  \label{eq:h1g}
   h_{a,{gg}}^{(1)}(\vec{k}) & = \frac{(2\pi)^{d/2}}{2 p_a^+}
  \int d k^- \frac{\overline{|{\cal M}^{(0)}_{g_ar^*\to gg}}|^2 }{2
    \vec{k}^2 \sqrt{ N_c^2-1} } d \Pi_2 \delta^{(d)} (p_a + k - p - q )
\notag \\
& 
=
 \int d z d^{2 + 2 \epsilon} {\vec k'} {F}_{ggg} ({\vec k}, {\vec k'}, z) h_{a,{\rm gluon}}^{(0)}(\vec{k'}) 
\end{align}
and
\begin{align}
  \label{eq:h1q}
   h_{a,{q{\bar q}}}^{(1)}(\vec{k}) & = \frac{(2\pi)^{d/2}}{2 p_a^+}
  \int d k^- \frac{\overline{|{\cal M}^{(0)}_{g_ar^*\to gg}}|^2 }{2
    \vec{k}^2 \sqrt{ N_c^2-1} } d \Pi_2 \delta^{(d)} (p_a + k - p - q )
\notag \\
& =
 \int d z d^{2 + 2 \epsilon} {\vec k'} {F}_{gq\bar{q}} ({\vec k}, {\vec k'}, z) h_{a,{\rm gluon}}^{(0)}(\vec{k'}) 
\end{align}
with 
\begin{align}
  \label{eq:Fggg}
   {F}_{ggg}({\vec k}, {\vec k'}, z)  & =  \frac{1}{2} \frac{\alpha_s}{2\pi \pi_\epsilon}{\cal P}_{gg}(z)\left[\frac{z^2\vec{k'}^2+(1-z)^2\vec{q}^2-z(1-z)\vec{q}\cdot \vec{k'} }{\vec{q}^2\vec{\Delta}^2}\right]
\end{align}
and 
\begin{align}
  \label{eq:Fgqqbar}
  {F}_{gq{\bar q}} ({\vec k}, {\vec k'}, z)  & =   \frac{\alpha_s}{2\pi \pi_\epsilon} n_f  {\cal P}_{qg}(z,\epsilon)       \frac{  1 }{{\vec q}^2} \left[\frac{C_f}{C_a}+z(1-z)\frac{\vec{q}\cdot \vec{\Delta}}{ \vec{\Delta}^2}\right],
\end{align}
where the overall factor 1/2 for  the $gg$ final state stems from the indistinguishability of identical bosons in the final state.\\
If we parametrize the momentum fraction $z$ in terms of the rapidity difference $\Delta y \equiv  y_p - q_q$ of the final state gluons, {\it i.e.}
\begin{align}
  \label{eq:zindeltaY}
  z & = \frac{e^{\Delta y}}{({\vec k'}^2/{\vec q}^2) + e^{\Delta y}    },
\end{align}
we can see that ${F}_{ggg}$ reduces in the
limits $\Delta y \to \pm \infty$ (including the corresponding Jacobian
factor) to half of the central production vertex of Eq.~(\ref{eq:defV}). To
regularize the resulting divergence of the rapidity integral we
introduce a lower bound $|\Delta y| > -\eta_b$, where $\eta_b$ is
again taken in the limit $\eta_b \to - \infty$, and define  $\mathcal{F}_{ggg}({\vec k}, {\vec k'}, z, \eta_b) ={F}_{ggg}({\vec k}, {\vec k'}, z)  \Theta( |\Delta y| + \eta_b) $. 
As in the case of virtual corrections it is now needed to subtract the contribution from gluon production at central rapidities to construct the complete differential cross section, schematically:
\begin{equation}
\parbox{3cm}{\center \includegraphics[width = 2.5cm]{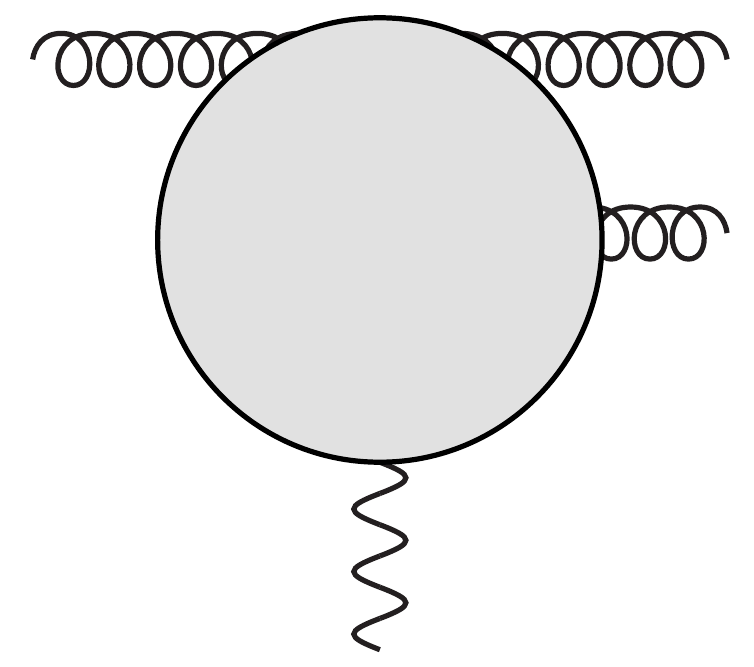}}
=
\parbox{3cm}{\center \includegraphics[width = 2.5cm]{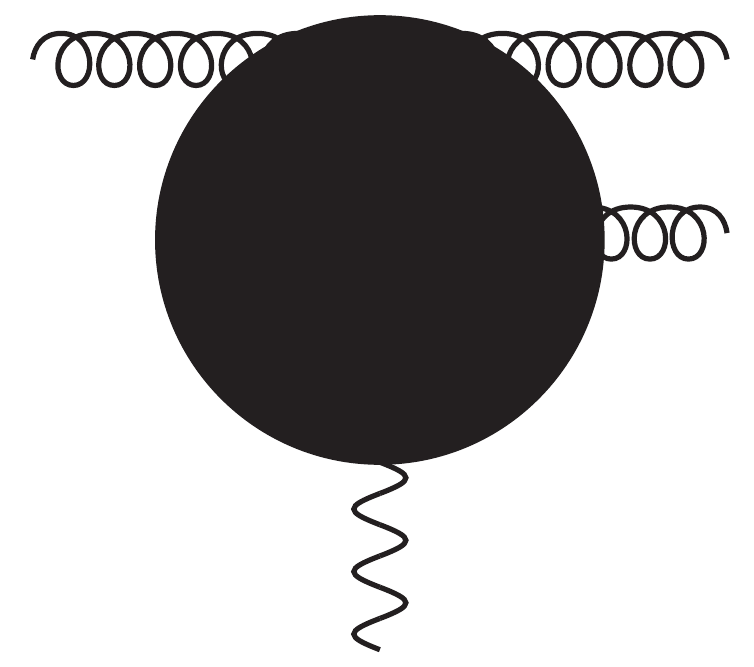}} -
\parbox{3cm}{\center \includegraphics[width = 2.5cm]{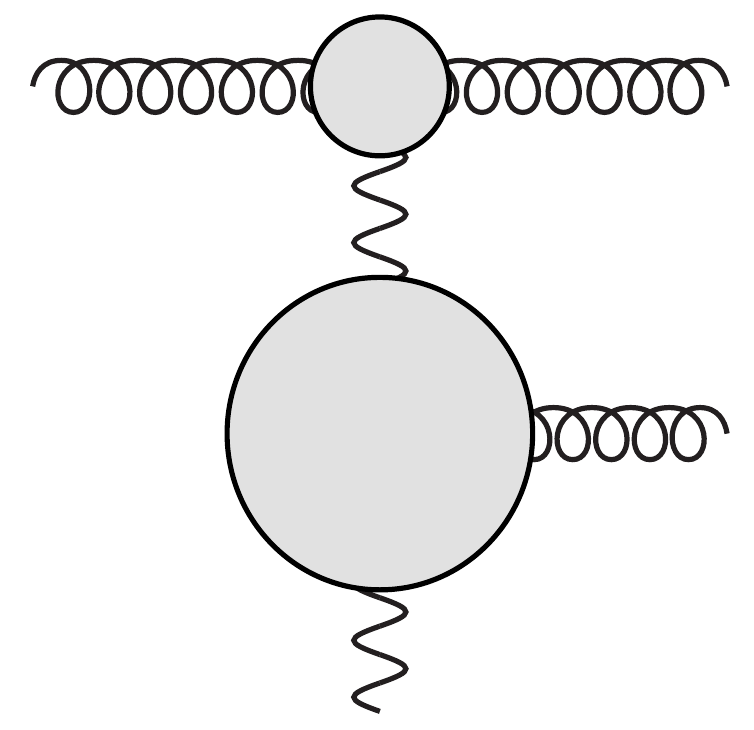}}
\end{equation}
This leads to the definition of the coefficient
\begin{align}
  \label{eq:realcoeff}
  \mathcal{G}^{(a)}_{ggg} ({\vec k}, {\vec k'}, z, \eta_b) & =  \mathcal{F}_{ggg} ({\vec k}, {\vec k'}, z, \eta_b)  -   \frac{1}{2}\left[ \frac{1}{z} 
 \mathcal{V} ({\vec k}, {\vec k'}; {\vec q}; \eta_a, \eta_b)  + \frac{1}{1-z}  \mathcal{V} ({\vec k}, {\vec q}; {\vec k'}; \eta_a, \eta_b) \right].
\end{align}
Defining finally the cross-section for quasi-elastic production as
\begin{align}
  \label{eq:prodquasiel}
  d\hat{\sigma}_{ab}^{(qea)} & =  h^{(0)}_{a, {\rm gluon}} ({\vec k'})  \mathcal{G}^{(a)}_{ggg} ({\vec k}, {\vec k'}, z, \eta_b) h^{(0)}_{b, {\rm gluon}} \, d z \,  d^{2 + 2\epsilon} {\vec k} \, d^{2 + 2\epsilon} {\vec k'},
\end{align}
the sum of central and quasi-elastic contributions,
\begin{equation}
d\hat{\sigma}_{ab}=d\hat{\sigma}_{ab}^{(c)}+d\hat{\sigma}_{ab}^{(qea)}+d\hat{\sigma}_{ab}^{(qeb)},
\end{equation}
turns out to be finite, if integrated over the gluon rapidity, with a
well-defined limit $\eta_{a,b} \to \pm \infty$. This completes our calculation of the gluon-initiated 
forward jet vertex at NLO using the high energy effective action constructed by Lipatov.

\section{Conclusions and Outlook}
\label{sec:concl}
We have performed the calculation of the forward jet vertex at
next-to-leading order in the BFKL formalism, offering an explicit
derivation of the gluon-initiated contribution.  This adds to the
previously calculated for the quark-initiated vertex and completes the
derivation of the full vertex. We have found agreement with previous
results in the literature. Our method of calculation is based on the
high energy effective action for QCD proposed by Lipatov, which is
proven to be very useful to streamline our calculations.  Our
subtraction and regularization procedure in order to avoid
over-counting of kinematic regions has been proven to work well in
both gluon and quark cases. We are now confident this method will help
in the calculation of further amplitudes at loop level.

\subsection*{Acknowledgements}

We acknowledge partial support from the European Comission under contract LHCPhenoNet (PITN-GA-2010-264564), 
the Comunidad de Madrid through Proyecto HEPHACOS ESP-1473, MICINN (FPA2010-17747), 
and Spanish MINECO's ``Centro de Excelencia Severo Ochoa" Programme under grant  SEV-2012-0249. M.H.  acknowledges support from the German Academic Exchange Service (DAAD), the U.S. Department of Energy under contract number DE-AC02-98CH10886 
and a BNL ``Laboratory Directed Research and Development" grant (LDRD 12-034)''. GC thanks the partial support from the Research Executive Agency (REA) 
of the European Union under the Grant Agreement number PIEF-GA-2011-298582, 
by MICINN (FPA2011-23778, FPA2007-60323 and CSD2007-00042 CPAN).

\appendix

\section{Appendix}
\label{sec:app}
In the following we present further details on the NLO calculation of both virtual and real corrections

\subsection{Virtual corrections}
\label{appvirt}
We refer for the notation to {Fig.}~\ref{fig:4}. Tadpole contributions vanish in dimensional regularization.

The contribution for each of these diagrams can be written in terms of master integrals labelled with the following notation: M,~S,~P, and Q denote, respectively, the existence of a propagator of the form $k^+(\to b\cdot k), k^2, (k-p_a)^2$ or $(k-q)^2$. The number at the end (0,\,1,\,2, or 3) indicates how many tensor indices are 
present in the numerator ({\it e.g.} 2 stands for a factor $k^\mu k^\nu$). $\xi=a^2=b^2=4e^{-\rho}$ are chosen to indicate the  squares of the new light-cone vectors. For diagrams (E), (F), (I) and (J)  in {Fig.}~\ref{fig:4} the contribution with reversed arrows is included. Diagrams (G) and (H) turn out to vanish completely. The 
symmetry factors for the diagrams, which are included, are equal to one apart from diagrams (C) and (D), for which it is two. In more detail, these are all the contributing 
expressions:
\begin{equation*}
\begin{aligned}
i{\cal M}_{\rm (A)}&=-\frac{ig^3}{2}\vec{q}^2f_{abc}\,N_c\int\frac{d^dk}{(2\pi)^d}\frac{(k^+-2p_a^+)^2\,\varepsilon\cdot\varepsilon^*+4\,\xi\,\varepsilon\cdot k\,\varepsilon^*\cdot(k-q)}{k^+k^2(k-p_a)^2(k-q)^2}\\
&=-\frac{ig^3}{2}\vec{q}^2f_{abc}\,N_c\,\{16e^{-\rho}\varepsilon_\mu\varepsilon^*_\nu {\rm [MSPQ2]}-16e^{-\rho}\varepsilon^*\cdot q\,\varepsilon_\mu{\rm[MSPQ1]}-4p_a^+\varepsilon\cdot\varepsilon^*{\rm [SPQ0]}\\
&+4(p_a^+)^2\varepsilon\cdot\varepsilon^*{\rm[MSPQ0]}+(n^+)_\mu\varepsilon\cdot\varepsilon^*{\rm [SPQ1]}\};\\\\
i{\cal M}_{\rm (B)}&=-\frac{ig^3}{2}f_{abc}\,N_c \int\frac{d^dk}{(2\pi)^d}\frac{1}{k^2(k-p_a)^2(k-q)^2}[4p_a^+\{\varepsilon\cdot\varepsilon^*((k-p_a)^2-\vec{q}^2)\\
&+4(\varepsilon\cdot k\,\varepsilon^*\cdot q-\varepsilon\cdot q\,\varepsilon^*\cdot k)\}+k^+\{7\vec{q}^2\varepsilon\cdot\varepsilon^*+(18+16\epsilon)\varepsilon\cdot k\,\varepsilon^*\cdot(k-q)\\ &+16\,\varepsilon\cdot q\,\varepsilon^*\cdot q\}]\\&=-\frac{ig^3}{2}f_{abc}\,N_c\{(18+16\epsilon)(n^+)_\mu\varepsilon_\nu\varepsilon^*_\rho{\rm [SPQ3]}-(18+16\epsilon)(n^+)_\mu\varepsilon_\nu\,\varepsilon^*\cdot q{\rm [SPQ2]}
\\&+[(n^+)_\mu(16\varepsilon\cdot q\,\varepsilon^*\cdot q+7\vec{q}^2\varepsilon\cdot\varepsilon^*)-16p_a^+(\varepsilon\cdot q\,\varepsilon^*_\mu-\varepsilon^*\cdot q\,\varepsilon_\mu)]{\rm [SPQ1]}\\
&-4p_a^+\vec{q}^2\varepsilon\cdot\varepsilon^*{\rm [SPQ0]}+4p_a^+\varepsilon\cdot\varepsilon^*{\rm [SQ0]}\};\\\\
\end{aligned}
\end{equation*}
\begin{equation}
\begin{aligned}
i{\cal M}_{\rm (C)}&=-\frac{ig^3}{2}f_{abc}\,N_c\int\frac{d^dk}{(2\pi)^d}\frac{1}{k^+k^2(k-q)^2}[4\xi(\varepsilon\cdot k\,\varepsilon^*\cdot q-\varepsilon\cdot q\,\varepsilon^*\cdot k)\\ &+\varepsilon\cdot\varepsilon^*(4k^+p_a^++\xi(2(k-p_a)^2-\vec{q}^2)]\\
&=-\frac{ig^3}{2}f_{abc}\,N_c\{-16e^{-\rho}(\varepsilon\cdot\varepsilon^*p_{a\mu}+\varepsilon\cdot q\,\varepsilon^*_\mu-\varepsilon^*\cdot q\,\varepsilon_\mu){\rm [MSQ1]}\\&-4e^{-\rho}\vec{q}^2\varepsilon\cdot\varepsilon^*{\rm [MSQ0]}+4p_a^+\varepsilon\cdot\varepsilon^*{\rm [SQ0]}\};\\\\
i{\cal M}_{\rm (D)}&=\frac{ig^3}{2\vec{q}^2} f_{abc}\,N_c\int\frac{d^dk}{(2\pi)^d}\frac{1}{k^2(k-q)^2}[-8p_a^+\vec{q}^2\varepsilon\cdot\varepsilon^*+(5+4\epsilon)k^+\{\varepsilon\cdot\varepsilon^*(\vec{q}^2\\&-2(k-p_a)^2)+4\,\varepsilon\cdot q\,\varepsilon^*\cdot k-4\,\varepsilon\cdot k\,\varepsilon^*\cdot q\}]\\&=\frac{ig^3}{2\vec{q}^2} f_{abc}\,N_c\{(20+16\epsilon)(n^+)_\mu(\varepsilon\cdot q\,\varepsilon^*_\nu-\varepsilon^*\cdot q\,\varepsilon_\nu+\varepsilon\cdot\varepsilon^*p_{a\nu}){\rm [SQ2]}\\&+(5+4\epsilon)\vec{q}^2\varepsilon\cdot\varepsilon^*(n^+)_\mu{\rm [SQ1]}-8p_a^+\vec{q}^2\varepsilon\cdot\varepsilon^*{\rm [SQ0]}\};\\\\
i{\cal M}_{\rm (E)}&=\frac{2ig^3}{\vec{q}^2} f_{abc}\,n_f\int\frac{d^dk}{(2\pi)^d}\frac{1}{k^2(k-q)^2}[p_a^+\vec{q}^2\varepsilon\cdot\varepsilon^*+k^+[\varepsilon\cdot\varepsilon^*(2(k-p_a)^2-\vec{q}^2)\\&+4(\varepsilon\cdot k\,\varepsilon^*\cdot q-\varepsilon\cdot q\,\varepsilon^*\cdot k)]]\\
&=-\frac{2ig^3}{\vec{q}^2}f_{abc}\,n_f\{4(n^+)_\mu(\varepsilon\cdot q\,\varepsilon^*_\nu-\varepsilon^*\cdot q\,\varepsilon_\nu+\varepsilon\cdot\varepsilon^*p_{a\nu}){\rm [SQ2]}\\&+\varepsilon\cdot\varepsilon^*\vec{q}^2(n^+)_\mu{\rm [SQ1]}-p_a^+\vec{q}^2\varepsilon\cdot\varepsilon^*{\rm [SQ0]}\};\\\\
i{\cal M}_{\rm (F)}&=\frac{ig^3}{2\vec{q}^2} f_{abc}\, N_c\int\frac{d^dk}{(2\pi)^d}\frac{k^+}{k^2(k-q)^2}[\varepsilon\cdot \varepsilon^*(2(k-p_a)^2-\vec{q}^2)\\&+4(\varepsilon\cdot k\,\varepsilon^*\cdot q-\varepsilon\cdot q\,\varepsilon^*\cdot k)]\\
&=-\frac{ig^3}{2\vec{q}^2}f_{abc}\, N_c\{4(n^+)_\mu(\varepsilon\cdot\varepsilon^* p_{a\nu}+\varepsilon\cdot q\,\varepsilon^*_\nu-\varepsilon^*\cdot q\,\varepsilon_\nu){\rm [SQ2]}\\&+\vec{q}^2\varepsilon\cdot\varepsilon^*(n^+)_\mu{\rm [SQ1]}\};\\\\
i{\cal M}_{\rm (I)}&=ig^3 f_{abc}\,n_f\int\frac{d^dk}{(2\pi)^d}\frac{1}{k^2(k-p_a)^2(k-q)^2}[p_a^+(-\vec{q}^2\varepsilon\cdot\varepsilon^*\\&+2(\varepsilon\cdot k\,\varepsilon^*\cdot q-\varepsilon\cdot q\,\varepsilon^*\cdot k))+k^+(\varepsilon\cdot\varepsilon^*(\vec{q}^2-2(k-p_a)^2)\\&+8\,\varepsilon\cdot k\,\varepsilon^*\cdot (k-q)+2\varepsilon\cdot q\,\varepsilon^*\cdot q)]\\&=ig^3f_{abc}\,n_f\,\{8(n^+)_\mu \varepsilon_\nu\,\varepsilon^*_\rho{\rm [SPQ3]}-8\varepsilon^*\cdot q(n^+)_\mu\varepsilon_\nu{\rm [SPQ2]}\\
&+[(\vec{q}^2\varepsilon\cdot\varepsilon^*+2\varepsilon\cdot q\,\varepsilon^*\cdot q)(n^+)_\mu+2p_a^+\varepsilon^*\cdot q\,\varepsilon_\mu-2p_a^+\varepsilon\cdot q\,\varepsilon^*_\mu]{\rm [SPQ1]}\\&-\vec{q}^2p_a^+\varepsilon\cdot \varepsilon^*{\rm [SPQ0]}\};\\\\
i{\cal M}_{\rm (J)}&=ig^3 f_{abc}\,N_c\int\frac{d^dk}{(2\pi)^d}\frac{k^+}{k^2(k-p_a)^2(k-q)^2}\varepsilon\cdot k\,\varepsilon^*\cdot(k-q)\\&=ig^3 f_{abc}\,N_c(n^+)_\mu\{\varepsilon_\nu\varepsilon^*_\rho{\rm [SPQ3]}-\varepsilon^*\cdot q\,\varepsilon_\nu{\rm [SPQ2]}\}. \nonumber
\end{aligned}
\end{equation}
Those integrals which are not suppressed in the $\rho\to\infty$ limit are:
\begin{equation}
\begin{aligned}
\left[{\rm SQ}0\right]&=\frac{i}{(4\pi)^{2+\epsilon}}(\vec{q}^2)^\epsilon\frac{\Gamma(-\epsilon)\Gamma^2(1+\epsilon)}{\Gamma(2+2\epsilon)};\qquad\\
\left[{\rm SQ}2\right]&=(g^{\mu\nu}\vec{q}^2+q^\mu q^\nu(4+2\epsilon))\frac{1}{4(3+2\epsilon)}\left[{\rm SQ}0\right];\\\left[{\rm SPQ}0\right]&=\frac{1+2\epsilon}{\epsilon\vec{q}^2} \left[{\rm SQ}0\right];\qquad\\ 
\left[{\rm SPQ}1\right]&=\left(q^\mu+\frac{1}{\epsilon}p_a^\mu\right)\left[{\rm SQ}0\right];\\ \left[{\rm SPQ}2\right]&=\left\{\frac{1}{2+2\epsilon}\left[\frac{1}{2}g^{\mu\nu}+\left(\frac{q^\mu p_a^\nu+p_a^\mu q^\nu}{\vec{q}^2}\right)+\frac{2}{\epsilon}\frac{p_a^\mu p_a^\nu}{\vec{q}^2}\right]+\frac{1}{\vec{q}^2}q^\mu q^\nu\right\}\left[{\rm SQ}0\right];\qquad\\
\left[{\rm SPQ}3\right]&=\frac{1}{\epsilon(1+\epsilon)(3+2\epsilon)}\bigg\{\frac{1}{\vec{q}^2}\bigg[p_a^\mu p_a^\nu p_a^\rho+\epsilon q^\mu p^\nu p^\rho+\frac{1}{2}\epsilon(1+\epsilon)q^\mu q^\nu p^\rho\\&-\frac{1}{6}(1-\epsilon)(2+\epsilon)^2 q^\mu q^\nu q^\rho\bigg]+\frac{\epsilon}{4}\left[p^\mu g^{\nu\rho}+(1+\epsilon)q^\mu g^{\nu\rho}\right]\\&+\text{\small{cyclic permutations of $\mu, \nu$ and $\rho$}}\bigg\}\left[{\rm SQ}0\right];\\\left[{\rm MSQ}1\right]&=\frac{b^\mu}{\xi}\left[{\rm SQ}0\right]+\frac{1}{2}q^\mu\left\{\frac{-i}{(4\pi)^{2+\epsilon}}\frac{\Gamma^2\left(\frac{1}{2}+\epsilon\right)\Gamma\left(\frac{1}{2}-\epsilon\right)\Gamma\left(\frac{1}{2}\right)}{\Gamma(1+2\epsilon)(\vec{q}^2)^{\frac{1}{2}-\epsilon}\xi^\frac{1}{2}}\right\};\\
\left[{\rm MSPQ}0\right]&=-\frac{i}{(4\pi)^{2+\epsilon}}\frac{(\vec{q}^2)^{\epsilon-1}}{p_a^+}\frac{\Gamma(1-\epsilon)\Gamma^2(\epsilon)}{\Gamma(2\epsilon)}\left(\ln\left[\frac{p_a^+}{|\vec{q}|}\right]+\frac{\rho}{2}+\frac{\psi(1)-2\psi(\epsilon)+\psi(1-\epsilon)}{2}\right). \nonumber
\end{aligned}
\end{equation}

\subsection{Details on the real corrections}
\label{sec:appreal}

We refer to Figs.~\ref{fig:6} and \ref{fig:7} for our notation. 
 We  have
 \begin{equation}
\label{os}
\begin{aligned}
 p^- & =\frac{(\vec{k}-\vec{q})^2}{(1-z)p_a^+},
&
 k^- &=\frac{(\vec{q}-z\vec{k})^2+z(1-z)\vec{k}^2}{z(1-z)p_a^+}.
\end{aligned}
\end{equation} 
Within the given choice of polarization vectors, diagrams (a), (b),
(c), (d) and (i) are immediately zero, while diagram (h) turns out to
vanish as well. The amplitudes for the non-vanishing diagrams can be written in the
following form
\begin{equation*}
\begin{aligned}
i{\cal M}_{\rm (e)}&=\varepsilon_{a\mu}\varepsilon^*_{b\nu}\varepsilon^*_{c\rho}2ig^2\frac{f_{ade}f_{bce}}{s}p_a^+\bigg\{g^{\nu\rho}[k^\mu(1-2z)-p^\mu+q^\mu]+g^{\mu\nu}(2p^\rho+q^\rho)\\&-g^{\mu\rho}(2q^\nu+p^\nu)\bigg\};\\
i{\cal M}_{\rm (f)}&=\varepsilon_{a\mu}\varepsilon^*_{b\nu}\varepsilon^*_{c\rho}(-ig^2)\frac{f_{abe}f_{cde}}{t}p_a^+\bigg\{-4z(g^{\nu\rho}p^\mu+g^{\mu\rho}p_a^\nu)+g^{\mu\nu}[k^\rho(2-z)+p^\rho(2+z)\\&+p_a^\rho(-2+3z)]\bigg\};\\
\end{aligned}
\end{equation*}
\begin{equation}
\begin{aligned}
i{\cal M}_{\rm (g)}&=\varepsilon_{a\mu}\varepsilon^*_{b\nu}\varepsilon^*_{c\rho}(-ig^2)\frac{f_{ace}f_{bde}}{u}p_a^+\bigg\{-4(1-z)[g^{\mu\nu}p_a^\rho+g^{\nu\rho}q^\mu]+g^{\mu\rho}[k^\nu(1+z)\\&+p_a^\nu(1-3z)+q^\nu(3-z)]\bigg\};\\
i{\cal M}_{\rm (j)}&=-\frac{2g^2f_{acd}\,t^c}{s}\varepsilon_{a\mu}[k^\mu(n^+)^\sigma-p_a^+g^{\mu\sigma}]\bar{u}(p)\gamma_\sigma v(q);\\
i{\cal M}_{\rm (k)}&=-\frac{ig^2\,t^dt^a}{u}\varepsilon_{a\mu}\bar{u}(p)\slashed{n}^+(\slashed{p}_a-\slashed{q})\gamma^\mu v(q);\\
i{\cal M}_{\rm (l)}&=\frac{ig^2\,t^at^d}{t}\varepsilon_{a\mu}\bar{u}(p)\gamma^\mu(\slashed{p}_a-\slashed{p})\slashed{n}^+ v(q).
\end{aligned}
\end{equation}

\end{document}